\documentclass[showpacs,amsmath,amssymb,aps,prb,10pt,
reprint,superscriptaddress]{revtex4-2}
\usepackage{bm}
\usepackage[breaklinks=true,colorlinks=true,linkcolor=blue,urlcolor=blue,citecolor=blue]{hyperref}
\usepackage{dcolumn}
\usepackage{amsmath,amssymb}
\usepackage{mathdots}
\usepackage{natbib}
\usepackage{soul,color}
\usepackage{graphicx}
\usepackage[note-name]{notes2bib}
\usepackage{mathptmx}
\usepackage[margin=1in]{geometry}

\begin{document}


\title{AC-flux-driven SQUID diode spectroscopy as a probe of
current-phase relations}

\author{Yuriy Yerin}
\affiliation{Istituto di Struttura della Materia of the National Research Council, via Salaria Km 29.3, I-00016 Monterotondo Stazione, Italy}
\affiliation{Department of Computer Engineering and Center of
Excellence of Superconductivity Research, Ankara University,
Ankara, 06100, Türkiye}
\affiliation{Cryogenic Quantum Electronics, EMG and LENA, Technische Universität Braunschweig, 38106, Braunschweig, Germany}

\author{Iman Askerzade}
\affiliation{Department of Computer Engineering and Center of
Excellence of Superconductivity Research, Ankara University,
Ankara, 06100, Türkiye}
\affiliation{Institute of Physics, H.~Cavid 33, Baku, AZ1143,
Azerbaijan}
\affiliation{Center for Theoretical Physics, Khazar University, 41 Mehseti Street, Baku, AZ1096, Azerbaijan}

\author{Alexey Fedorchenko}
\affiliation{B. Verkin Institute for Low Temperature Physics and Engineering of the National Academy of Sciences of Ukraine, Kharkiv 61103, Ukraine}

\author{Ali Gencer}
\affiliation{Ankara University, Superconductor Technologies Application and Research Center, Ankara, Türkiye}
\affiliation{Science Faculty, Physics Department, Ankara University, Ankara, Türkiye}

\author{Oleksandr Dobrovolskiy}
\affiliation{Cryogenic Quantum Electronics, EMG and LENA, Technische Universität Braunschweig, 38106, Braunschweig, Germany}
\affiliation{FLUXONICS---The European Foundry for Superconducting Electronics e.V., 38116 Braunschweig, Germany}

\date{\today}

\begin{abstract}
The current–phase relation (CPR) of a Josephson junction encodes microscopic information on superconducting states through higher-order and fractional harmonics. 
However, their unambiguous extraction is challenging, as different CPR components produce nearly identical static interference patterns that are further obscured by device asymmetries, damping, and dynamical effects. 
Here, we propose probing individual CPR harmonics via the ac magnetic-flux-driven diode effect in asymmetric dc SQUIDs with unequal junction critical currents. 
Using two complementary reductions of the fast-driven dynamics---a Kapitza-type perturbation theory for the conventional junction and a Jacobi–Anger averaging for a general CPR---we show that ac flux modulation dresses each harmonic with a distinct Bessel function, yielding characteristic signatures in the diode efficiency $\eta(\phi_{\rm ac},\omega)$ as a function of ac flux amplitude $\phi_{\rm ac}$ and frequency $\omega$. 
We verify and extend these predictions by numerical solutions of the coupled dynamical equations for CPRs containing $\sin\varphi$, $\sin(\varphi/2)$, and $\sin 2\varphi$ terms ($\varphi$: superconducting phase difference), and construct phase diagrams of $\eta(\phi_{\rm ac},\omega)$. 
Distinct CPR components are revealed to produce characteristic weak, sparse, dense, or intermodulated arc patterns that remain robust in both overdamped and underdamped regimes. 
This suggests ac-flux-driven SQUID diode spectroscopy as a probe of current–phase relations in topological materials, multiband systems, and other unconventional superconductors.
\end{abstract}

\maketitle

\section{Introduction}
\label{sec:intro}

The superconducting diode effect, characterized by unequal
critical currents for positive and negative current bias,
$I_\mathrm{c}^+ \neq |I_\mathrm{c}^-|$, has emerged as a key manifestation of nonreciprocal transport in Josephson systems~\cite{Nadeem2023,review2025,InglaAynes2025}. Unlike conventional semiconductor diodes, a superconducting diode supports supercurrent in one direction while suppressing it in the opposite direction, enabling energy-efficient cryogenic rectifiers, on-chip circulators, and nonvolatile memory elements~\cite{GolodNatComms2022,InglaAynes2025}. The effect has been experimentally observed across a wide range of systems, including superconductors with broken inversion symmetry~\cite{WakatsukiSciAdv2017,Zhang2020,Itahashi20,Lyu2021,Narita2022,Ghosh2024,Soori2024,Soori2025}, van der Waals heterostructures~\cite{wu22,lin22,margineda2026}, proximitized semiconductor nanowires and quantum dots~\cite{Turini22,PalNatPhys2022,Debnath2024,Debnath2025}, and Josephson junction arrays~\cite{GolodNatComms2022,Wei2025arXivHTSdiode}.

Microscopically, the superconducting diode effect arises from broken time-reversal and/or inversion (chiral) symmetry~\cite{Tokura2018,Nadeem2023}. In Josephson devices, time-reversal is broken by external magnetic flux and inversion by junction or circuit asymmetry. This general framework encompasses a wide variety of physical mechanisms:
spin-orbit coupling in Rashba systems~\cite{Ilic22,Edelstein95,Edelstein_1996}, finite-momentum Cooper pairing~\cite{yuan22,AGTERBERG200313,agterberg2011},
magnetic proximity effects~\cite{Narita2022,Buzdin,roig2024},
and nontrivial topology of the superconducting
condensate~\cite{Zinkl,BoLu2022,BoLu2023,Cayao_Majorana, Yerin_fractal}. Phenomenological and microscopic descriptions of these symmetry-breaking mechanisms have been developed~\cite{he22,dai22,Hasan2024SDEHelical,Cadorim2024PRApp,sun23}, including in multiband \cite{Yerin_diode,Yerin_SQUID,Yerin2014,Yerin_review,Kiyko} and, more specifically, multicomponent superconductors, where the interband phase difference provides an additional degree of freedom~\cite{Tanaka_review}.

The current–phase relation (CPR) of a Josephson junction reflects the microscopic superconducting state and offers a route to realizing the superconducting diode effect. In a conventional tunnel junction, CPR is purely sinusoidal, with the supercurrent $I_\mathrm{s}(\varphi) = I_\mathrm{c}\sin\varphi$, where $I_\mathrm{c}$ is the critical current and $\varphi$ the phase difference across the junction. However, in several classes of unconventional junctions, CPR acquires additional subharmonic and higher-harmonic components~\cite{askerzade_2012,ModernAspects2017,askerzade_2015}. For instance, high-transparency junctions develop a second-harmonic contribution $\sin 2\varphi$, whose amplitude increases with junction transparency~\cite{Golubov_review,askerzade_2015}. Topological Josephson junctions between a conventional superconductor and a topological material hosting Majorana zero modes exhibit a $4\pi$-periodic current–phase relation with a dominant $\sin(\varphi/2)$ component, known as the fractional Josephson effect~\cite{Cayao_Majorana,mondal2025}. This $4\pi$-periodicity is a direct consequence of the non-Abelian
statistics of Majorana fermions and has been proposed as a transport-level signature of topological
superconductivity~\cite{Cayao_Majorana,BoLu2022,BoLu2023}.
Experimentally, this signature manifests as missing odd Shapiro steps under microwave irradiation and has been reported in HgTe-based topological junctions \cite{Wiedenmann2016}, InAs quantum wells \cite{Dartiailh2021}, and
topological insulator devices \cite{Rosen2024, Wu2025, Frolov2025}. Notably, the fractional Josephson effect is not exclusive to topological systems but can also arise in high-transparency conventional junctions via Landau--Zener transitions between Andreev bound states~\cite{mondal2025,Cuozzo2024}.

The enhanced flux sensitivity of superconducting quantum interference devices (SQUIDs) makes them natural interferometric extensions of Josephson junctions, with the junction CPR affecting the SQUID diode efficiency. Thus, in asymmetric SQUIDs formed by one conventional and one topological junction, the $4\pi$-periodic channel yields a diode effect with tunable polarity via external flux or junction parameters and enhanced efficiency compared to sinusoidal asymmetric SQUIDs over a broad parameter range~\cite{Cuozzo2024,Cayao_Majorana}. At the same time, a key challenge is the inverse problem of reconstructing the CPR spectrum from the measured diode response. 

Existing approaches to this problem rely on flux-dependent dc transport measurements and microwave spectroscopy techniques such as Shapiro-step analysis~\cite{Wu2025,Matsuo2025}. The dc approach typically probes the flux dependence of $I_\mathrm{c}^{\pm}$ via asymmetries in the SQUID's Fraunhofer-like oscillation pattern. However, different CPR components can yield qualitatively similar flux-dependent behavior, hindering their unambiguous separation. Shapiro step analysis is more sensitive to CPR content but is susceptible to microwave transmission-line effects that may distort the observed harmonic structure. Recently, it was shown that microwave power at fixed frequency can tune the diode polarity in asymmetric SQUIDs with a topological junction, with a characteristic difference in switching power between topological and conventional cases~\cite{Cuozzo2024, yerin2026kapitza}. While this provides a useful discriminator, previous studies have typically varied the microwave power at fixed frequency, leaving the two-dimensional $(\phi_\mathrm{ac}, \omega)$ parameter space unexplored. Here, $\phi_\mathrm{ac}$ and $\omega$ denote, respectively, the amplitude and frequency of the applied ac magnetic flux.

In this work, we propose ac flux modulation as a spectroscopic probe of CPRs in asymmetric dc SQUIDs. In contrast to fixed-frequency approaches, we characterize the diode efficiency $\eta(\phi_\mathrm{ac}, \omega)$ over the full two-dimensional parameter space by sweeping both the amplitude and frequency of the applied ac flux. We show, both analytically and numerically, that this two-dimensional map encodes distinct and identifiable fingerprints of the CPR components of the junctions. The discussed results are relevant for probing CPRs in junctions based on topological insulators, multiband superconductors, and proximitized heterostructures.

The paper is organized as follows. Section~\ref{sec:formalism} introduces the model and equations of motion and summarizes the main analytical results. Section~\ref{sec:static} presents the static and quasi-dynamical diode response and its dependence on the driving parameters. Section~\ref{sec:numerics} presents the two-dimensional $\eta(\phi_\mathrm{ac},\omega)$ phase diagrams for representative CPR cases, which are then discussed in Sec.~\ref{sec:diag_summary}. We outline a possible experimental realization in Sec.~\ref{sec:experiment} and conclude in Sec.~\ref{sec:conclusions}. Detailed analytical derivations are provided in Supplemental Material~\cite{SupplementalMaterial}.

\section{Model and analytical results}
\label{sec:formalism}

\subsection{Main equations}

\begin{figure}[t!]
    \centering
    \includegraphics[width=1\linewidth]{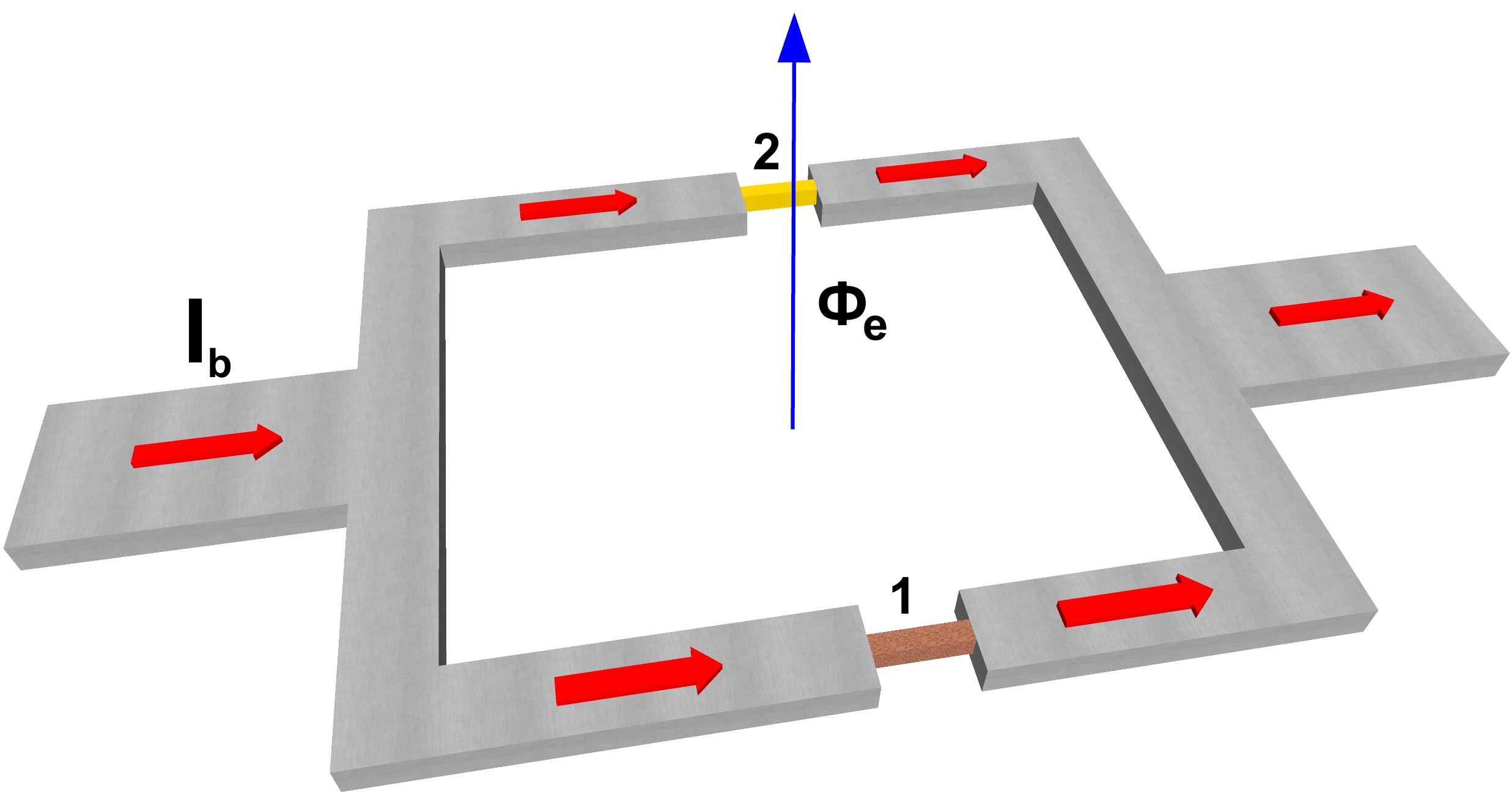}
    \caption{Schematic of the asymmetric dc SQUID geometry. The device consists of two Josephson junctions (1 and 2) threaded by an external magnetic flux $\Phi_\mathrm{e} = \Phi_\mathrm{dc} + \Phi_\mathrm{ac}\cos\omega t$, with dc and ac flux components and biased by the dc current $I_\mathrm{b}$. Junction 1 is a conventional junction with a purely sinusoidal CPR, while junction 2 may support subharmonic and second-harmonic contributions, as described by Eq.~\eqref{eq:CPR2}.}
    \label{fig:squid_sketch}
\end{figure}

We consider a dc SQUID consisting of two Josephson junctions (Fig.~\ref{fig:squid_sketch}) with CPRs $I_{\mathrm{s},k}$ ($k=1,2$). The junctions form a superconducting loop of geometric inductance $L$ threaded by an external magnetic flux $\Phi_\mathrm{e}=\Phi_\mathrm{dc}+\Phi_\mathrm{ac}\cos\omega t$ and biased by a dc current $I_\mathrm{b}$. Each junction is modeled within the resistively and capacitively shunted junction (RCSJ) framework, with identical $R$ and $C$ assumed for simplicity. The gauge-invariant phase differences $\varphi_1$ and $\varphi_2$ then obey~\cite{BaronePaterno1982, Waal1984}
\begin{align}
\frac{\hbar C}{2e}\,\ddot{\varphi}_1
+\frac{\hbar}{2eR}\,\dot{\varphi}_1
+I_{\mathrm{s},1}(\varphi_1)
&=\frac{I_\mathrm{b}}{2}-J,
\label{eq:EOM_dim1}\\
\frac{\hbar C}{2e}\,\ddot{\varphi}_2
+\frac{\hbar}{2eR}\,\dot{\varphi}_2
+I_{\mathrm{s},2}(\varphi_2)
&=\frac{I_\mathrm{b}}{2}+J,
\label{eq:EOM_dim2}
\end{align}
where the inductive (loop) current is
\begin{equation}
J=\frac{1}{L}\!\left[
\frac{\hbar}{2e}\,(\varphi_1-\varphi_2)-\Phi_\mathrm{e}
\right].
\label{eq:coupling_dim}
\end{equation}

It is convenient to recast Eqs.~(\ref{eq:EOM_dim1})–(\ref{eq:coupling_dim}) in dimensionless form. Time is rescaled by the inverse plasma frequency $\tau_\mathrm{p}^{-1}=2eI_\mathrm{c1}^{(0)}R/\hbar$, currents by $I_\mathrm{c1}^{(0)}$, where $I_\mathrm{c1}^{(0)}$ is the reference critical current of the conventional Josephson junction, voltage by $I_\mathrm{c1}^{(0)}R$, and flux by $\Phi_0=h/2e$, yielding the dimensionless variables $i_\mathrm{b}$, $j$, and $\phi_\mathrm{e}$. The equations of motion then take the standard RCSJ form
\begin{align}
\beta_\mathrm{c} \ddot{\varphi}_1 + \dot{\varphi}_1
+ i_\mathrm{c1}(\varphi_1) &= \frac{i_\mathrm{b}}{2} - j,
\label{eq:EOM1}\\
\beta_\mathrm{c} \ddot{\varphi}_2 + \dot{\varphi}_2
+ i_\mathrm{c2}(\varphi_2) &= \frac{i_\mathrm{b}}{2} + j,
\label{eq:EOM2}
\end{align}
where $i_{\mathrm{c}k}(\varphi_k)\equiv I_{\mathrm{s},k}(\varphi_k)/I_\mathrm{c1}^{(0)}$ is the CPR of junctions.

We model $i_{\mathrm{c}1}(\varphi_1)$ as a conventional sinusoidal tunnel junction, while $i_{\mathrm{c}2}(\varphi_2)$ also includes subharmonic and second-harmonic contributions:
\begin{align}
i_\mathrm{c1}(\varphi_1) &= \sin\varphi_1,
\label{eq:CPR1}\\
i_\mathrm{c2}(\varphi_2) &= \alpha\sin\varphi_2
+ i_\mathrm{c2}^{(1)}\sin\!\left(\frac{\varphi_2}{2}\right)
+ i_\mathrm{c2}^{(2)}\sin 2\varphi_2.
\label{eq:CPR2}
\end{align}
The parameter $0 \le \alpha \le 1$ controls
the amplitude asymmetry between the two junctions. The two
additional terms in Eq.~(\ref{eq:CPR2}) account for non-sinusoidal (exotic) CPR contributions. Namely, the $\sin(\varphi/2)$ term is $4\pi$-periodic and corresponds to the fractional Josephson effect. 
The $\sin 2\varphi$ term is $\pi$-periodic and represents the leading second-harmonic correction arising from high-transparency channels in junctions described by the generalized Ambegaokar–Baratoff relation~\cite{Ambegaokar,Golubov_review}. Both terms break the conventional $2\pi$-periodic CPR and, as we show below, lead to qualitatively distinct fingerprints in the ac-flux-driven diode response.

The dimensionless inductive coupling current is
\begin{equation}
    j = \frac{\varphi_1 - \varphi_2 - 2\pi\phi_\mathrm{e}}{\pi\beta_\mathrm{L}},
    \label{eq:coupling}
\end{equation}
while the dimensionless parameters are
\begin{equation}
    \beta_\mathrm{L}=\frac{2LI_\mathrm{c1}^{(0)}}{\Phi_0},
    \qquad
    \beta_\mathrm{c}=\frac{2\pi I_\mathrm{c1}^{(0)}R^2 C}{\Phi_0},
    \label{eq:beta_def}
\end{equation}
corresponding to the screening parameter and the Stewart-McCumber parameter, respectively.

The total dimensionless external flux $\phi_\mathrm{e}$ is 
\begin{equation}
    \phi_\mathrm{e} = \phi_\mathrm{dc} + \phi_\mathrm{ac}\cos\omega t,
    \label{eq:flux}
\end{equation}
where $\phi_\mathrm{dc}$ is the dc flux bias breaking time-reversal
symmetry and $\phi_\mathrm{ac}\cos\omega t$ is the ac flux component of tunable amplitude $\phi_\mathrm{ac}$ and
frequency $\omega$. The latter is in units of $\omega_\mathrm{p} = \tau_\mathrm{p}^{-1}$, where $\omega_\mathrm{p}$ is the plasma frequency.

The normalized voltage across the SQUID and its time average are
\begin{equation}
    v = \frac{\dot{\varphi}_1+\dot{\varphi}_2}{2},
    \qquad
    \langle v\rangle = \frac{1}{T}\int_0^T v(t)\,dt,
    \label{eq:voltage}
\end{equation}
where the dot denotes derivative with
respect to dimensionless time and $T$ is taken much longer than all dynamical timescales of the system. 

The diode efficiency is defined as
\begin{equation}
\eta = \frac{I_\mathrm{c}^+ - |I_\mathrm{c}^-|}{I_\mathrm{c}^+ + |I_\mathrm{c}^-|},
\label{eq:eta_def}
\end{equation}
where $I_\mathrm{c}^\pm$ are the positive and negative switching critical currents extracted from the current-voltage curve.

\subsection{Analytical results}
\label{sec:analytics_summary}

For pure sinusoidal junctions with amplitude asymmetry ($i_\mathrm{c2}^{(1)}=i_\mathrm{c2}^{(2)}=0$), we develop a Kapitza-type perturbation theory in the limit $\beta_\mathrm{L}\ll 1$ and $\xi_0\ll 1$. Here, $\xi_0$ is the amplitude of the fast difference-phase
oscillation driven by the ac flux. The main result \cite{SupplementalMaterial} for the diode efficiency is
\begin{equation}
    \eta \simeq \eta_\mathrm{L} + \Delta\eta_\mathrm{ac},
    \label{eq:eta_total}
\end{equation}
where the static finite-inductance contribution is:
\begin{equation}
    \eta_\mathrm{L} = \frac{\pi\beta_\mathrm{L}\,\alpha(1-\alpha^2)
    \sin(2\pi\phi_\mathrm{dc})}{16R_0^3},
    \label{eq:eta_L}
\end{equation}
and the ac-induced correction is given by an expression:
\begin{equation}
    \Delta\eta_\mathrm{ac} =
    \frac{\alpha(1-\alpha^2)\,\phi_\mathrm{ac}^2\,\sin(2\pi\phi_\mathrm{dc})}
    {4\beta_\mathrm{c}\beta_\mathrm{L}^2\,R_0^3\,
    (1+\beta_\mathrm{c}^2\omega^2)\,
    [(\omega_\mathrm{L}^2-\omega^2)^2+\gamma^2\omega^2]},
    \label{eq:eta_ac}
\end{equation}
with
\begin{equation}
    R_0=\sqrt{
    \!\left(\frac{1+\alpha}{2}\right)^{\!2}\!\cos^2\!\pi\phi_\mathrm{dc}
    +\left(\frac{1-\alpha}{2}\right)^{\!2}\!\sin^2\!\pi\phi_\mathrm{dc}},
    \label{eq:R0}
\end{equation}
\begin{equation}
    \omega_\mathrm{L}^2=\frac{2}{\pi\beta_\mathrm{c}\beta_\mathrm{L}},
    \qquad
    \gamma=\frac{1}{\beta_\mathrm{c}}.
    \label{eq:omegaL}
\end{equation}

The amplitude $\xi_0$ of the fast difference-phase
oscillation introduced above is set by the driven dynamics
of the difference mode. Linearizing the difference-phase
equation about the inductively pinned difference-phase value
$\varphi_{-} =(\varphi_1-\varphi_2)/2\simeq\pi\phi_\mathrm{dc}$, the fast component obeys a
driven damped-oscillator equation
[Eq.~(\ref{eq:fast_osc}) of the Supplemental Material] whose
steady-state amplitude is
\begin{equation}
    \xi_0=
    \frac{2\phi_\mathrm{ac}/(\beta_\mathrm{c}\beta_\mathrm{L})}
    {\sqrt{(\omega_\mathrm{L}^2-\omega^2)^2+\gamma^2\omega^2}},
    \label{eq:xi0_main}
\end{equation}
with $\omega_\mathrm{L}$ and $\gamma$ given by Eq.~(\ref{eq:omegaL}). Thus, $\xi_0$ grows linearly with the ac drive amplitude
$\phi_\mathrm{ac}$ and is resonantly enhanced near $\omega=\omega_\mathrm{L}$. It is the single dynamical quantity through which the ac flux modulation enters the effective current-phase relation, and it controls all of the Bessel-function dressings discussed
in what follows.

The frequency $\omega_\mathrm{L}$ plays a central role in the
subsequent analysis and deserves a brief physical
interpretation. In our normalized units, 
$\omega_\mathrm{L}$ is the natural eigenfrequency of the
difference-phase mode $\varphi_{-} =(\varphi_1-\varphi_2)/2$ in the linearized limit where the Josephson nonlinearity is replaced by the inductive restoring force [Eq.~(\ref{eq:fast_osc}) of the Supplemental Material]. Equivalently, $\omega_L$ can be written as $\omega_L=1/\sqrt{LC_{\rm eff}}$ in dimensional
form, where $C_{\rm eff}=C/2$ is the effective
capacitance of the difference-phase mode loaded by both junctions
in parallel: $\omega_L$ is therefore the LC plasma
resonance of the SQUID loop, not the single-junction
plasma frequency $\tau_p^{-1}$.
Because $\omega_\mathrm{L}$ is the only intrinsic frequency scale appearing in the denominator of $\xi_0$ [see Eq.~(\ref{eq:xi0_main})], it controls the response of the entire SQUID to the ac drive: for $\omega\ll\omega_\mathrm{L}$ the system follows the ac flux quasi-statically; for $\omega\sim\omega_\mathrm{L}$ the
difference-mode oscillation is resonantly enhanced; and for $\omega\gg\omega_\mathrm{L}$ the ac drive averages out and the response approaches the adiabatic limit. All three regimes appear directly in the numerical phase diagrams of Sec.~\ref{sec:numerics}.

Several important consequences follow from Eqs.~(\ref{eq:eta_total})--(\ref{eq:eta_ac}). First, a finite loop inductance $\beta_\mathrm{L}$ already produces a nonzero $\eta_\mathrm{L}$ even at $\phi_\mathrm{ac}=0$. This static contribution arises because the finite inductance allows the difference phase to respond to the common phase, generating an effective second harmonic in CPR. Second, within their domain of validity ($\xi_0\ll 1$), Eqs.~(\ref{eq:eta_total})--(\ref{eq:eta_ac}) predict that the ac-induced correction $\Delta\eta_\mathrm{ac}$ grows smoothly as $\phi_\mathrm{ac}^2$. As we discuss below and in~\cite{SupplementalMaterial}, this monotonic growth is the small-argument limit of a Bessel-dressed expression that does develop oscillatory structure at larger drive amplitude. Third, the frequency dependence is controlled by two smooth response factors: $[(\omega_\mathrm{L}^2-\omega^2)^2+\gamma^2\omega^2]^{-1}$ from the  driven difference-mode oscillation, and $(1+\beta_\mathrm{c}^2\omega^2)^{-1}$ from the common-mode susceptibility at the switching point. Neither factor produces an oscillatory structure in $\omega$ by itself.

The three factors $\alpha(1-\alpha^2)$, $\sin(2\pi\phi_\mathrm{dc})$, and $\phi_\mathrm{ac}^2$ in the numerator of Eq.~(\ref{eq:eta_ac}) ensure that $\Delta\eta_\mathrm{ac}$ vanishes in three distinct symmetry-protected limits: for symmetric junctions ($\alpha=1$), at time-reversal symmetric flux values ($\phi_\mathrm{dc}=0$ or $\frac{1}{2}$), and in the absence of ac drive ($\phi_\mathrm{ac}=0$).

Eqs.~(\ref{eq:eta_L}) and~(\ref{eq:eta_ac}) are derived
under the assumption $\xi_0\ll 1$. Retaining the fast
oscillation $\xi(t)$ exactly inside the trigonometric
functions, rather than expanding to leading order, yields a
non-perturbative generalization of the induced diode
efficiency~\cite{SupplementalMaterial}
\begin{equation}
\eta \simeq \frac{K_{\mathrm{tot}}\, \bigl[1+J_0(2\xi_0)\bigr]\,
\alpha(1-\alpha^2)\sin(2\pi\phi_\mathrm{dc})}
{16\,J_0(\xi_0)\,R_0^3},
\label{eq:eta_a_nonpert}
\end{equation}
with
\begin{equation}
K_{\mathrm{tot}}
=
\frac{\pi\beta_\mathrm{L}}{2}
+
\frac{\beta_\mathrm{c}\,\xi_0^2}{2(1+\beta_\mathrm{c}^2\omega^2)}.
\label{eq:Ktot_main}
\end{equation}
In the limit $\xi_0\ll 1$ one has $J_0(2\xi_0)\to 1$ and
$J_0(\xi_0)\to 1$, so the bracket $[1+J_0(2\xi_0)]/[2J_0(\xi_0)]\to 1$
and Eq.~(\ref{eq:eta_a_nonpert}) reduces exactly to
$\eta_\mathrm{L}+\Delta\eta_\mathrm{ac}$ of Eqs.~(\ref{eq:eta_total})--(\ref{eq:eta_ac}).
At larger drive amplitude, however, the dressed first
harmonic $J_0(\xi_0)R_0$ in the denominator passes through
zero whenever $\xi_0$ reaches a zero of $J_0$, and the
leading-order reduction breaks down there. 

Eqs.~(\ref{eq:eta_L}) and~(\ref{eq:eta_ac}) capture the symmetry, sign, and leading perturbative
scaling of the diode response. However, they are not
uniformly valid near the lines where the effective first
harmonic $R_0$ becomes small, in particular for
$\phi_\mathrm{dc}\approx n+\tfrac{1}{2}$ and $\alpha\approx 1$, where the perturbative expansion in the effective second
harmonic breaks down. In this regime the inductive
correction becomes comparable to or larger than the leading first-harmonic amplitude $R_0$, so the small-$\beta_\mathrm{L}$ Taylor expansion underlying Eqs.~(\ref{eq:eta_L})--(\ref{eq:eta_ac}) is no longer controlled. Within the perturbative regime, however, the analytical formulas provide a reliable description of the diode response, as we verify in Sec.~\ref{sec:numerics}.

In the limit of strong inductive screening $\beta_\mathrm{L}\ll 1$, the difference phase $\varphi_-=(\varphi_1-\varphi_2)/2$ is slaved to the external flux, $\varphi_-(t)\simeq \pi\phi_\mathrm{dc}+\xi(t)$, where $\xi(t)$ is the small fast oscillation driven by the ac flux component ~\cite{SupplementalMaterial}. Substituting this decomposition into the common-phase equation of motion and averaging over one ac period using the Jacobi-Anger identity ~\cite{AS1964},
\begin{equation}
    \bigl\langle\sin\!\bigl(A+z\cos(\omega t-\delta)\bigr)
    \bigr\rangle
    =J_0(z)\sin A,
    \label{eq:jacobi_anger_main}
\end{equation}
one finds that each CPR harmonic acquires its own Bessel-function dressing factor~\cite{SupplementalMaterial}:
\begin{equation}
    \sin\varphi\to J_0(\xi_0),
    \quad
    \sin\!\displaystyle\frac{\varphi}{2}\to J_0\!\displaystyle\frac{\xi_0}{2},
    \quad
    \sin 2\varphi\to J_0(2\xi_0).
\label{eq:bessel_dressing_main}
\end{equation}
The argument of each Bessel function is set by the effective phase-modulation amplitude experienced by that harmonic.

Computing the resulting effective CPR
$I_{\mathrm{eff}}(\varphi_+)$ near its extrema and
evaluating Eq.~(\ref{eq:eta_def}) to leading order in the
subleading harmonics~\cite{SupplementalMaterial}, the diode efficiency in the fast-averaged pinned-flux
regime is
\begin{widetext}
\begin{equation}
\eta\simeq
-\frac{i_\mathrm{c2}^{(1)}\,J_0(\xi_0/2)}
{2\sqrt{2}\,J_0(\xi_0)\,R_0}
\sin\!\Bigl(\displaystyle\frac{\pi\phi_\mathrm{dc}+\theta}{2}\Bigr)
+
\frac{i_\mathrm{c2}^{(2)}\,J_0(2\xi_0)}
{2\,J_0(\xi_0)\,R_0}
\sin\!\Bigl(2(\pi\phi_\mathrm{dc}+\theta)\Bigr),
\label{eq:eta_jacobi_anger_main}
\end{equation}
\end{widetext}
where $\theta$ is the phase shift of
the dressed first harmonic
\begin{equation}
    \tan\theta =
    \frac{(1-\alpha)\sin\pi\phi_\mathrm{dc}}
    {(1+\alpha)\cos\pi\phi_\mathrm{dc}},
    \label{eq:theta_def}
\end{equation}
and $R_0$ is given by Eq.~(\ref{eq:R0}). Equation \eqref{eq:theta_def} yields three predictions for the location of the arc-like structures that we will observe in the numerical phase diagrams.

Firstly, in a dc SQUID with a pure $\sin(\varphi/2)$ subharmonic
($i_\mathrm{c2}^{(1)}\neq 0$, $i_\mathrm{c2}^{(2)}=0$), the dominant
nodal lines $\eta=0$ are governed by the zeros of
$J_0(\xi_0/2)$. Because the argument is halved, these
zeros are well separated in $\xi_0$, and the corresponding arcs in the $(\phi_\mathrm{ac},\omega)$ plane are widely spaced. Secondly, in a dc SQUID with a pure $\sin 2\varphi$ second harmonic ($i_\mathrm{c2}^{(1)}=0$, $i_\mathrm{c2}^{(2)}\neq 0$), the
nodal lines are governed by the zeros of $J_0(2\xi_0)$.
The argument is now doubled, and the arcs in the $(\phi_\mathrm{ac},\omega)$ plane are correspondingly denser, roughly by a factor of four with respect to the subharmonic case.
Finally, in the mixed case the two contributions in
Eq.~(\ref{eq:eta_jacobi_anger_main}) compete and can partially cancel each other, producing intermodulation features that lie at neither set of pure zeros.

For the conventional purely sinusoidal asymmetric case
($i_\mathrm{c2}^{(1)}=i_\mathrm{c2}^{(2)}=0$), Eq.~(\ref{eq:eta_jacobi_anger_main}) gives $\eta=0$ at this leading order. A nonzero diode response in this case requires the next-order finite-$\beta_\mathrm{L}$ feedback and dynamical back-action mechanisms analyzed in Sec.~\ref{sec:analytics_summary} and \cite{SupplementalMaterial}, which yield Eqs.~(\ref{eq:eta_total})--(\ref{eq:eta_ac}) at small $\xi_0$
and the non-perturbative expression Eq.~(\ref{eq:eta_a_nonpert}) more generally.

\section{Static and dynamical diode behavior}
\label{sec:static}

\begin{figure}[t]
    \centering
    \includegraphics[width=0.85\linewidth]{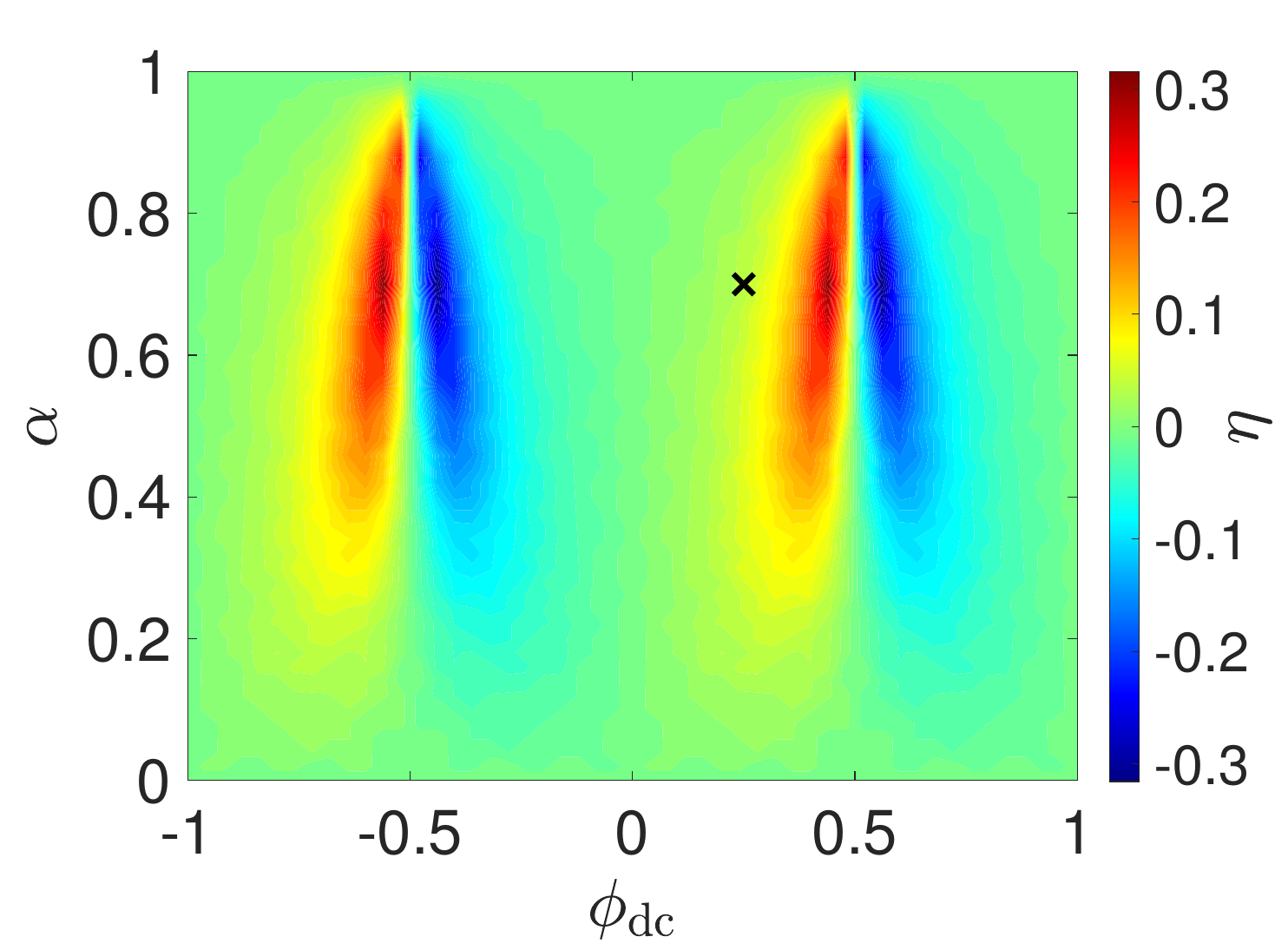}
    \caption{Diode efficiency $\eta(\phi_\mathrm{dc},\alpha)$ of a dc SQUID with two purely sinusoidal Josephson junctions for $\beta_\mathrm{c}=1$ and $\beta_\mathrm{L}=0.15$ in the static regime ($\phi_\mathrm{ac}=0$). The cross marks the working point ($\phi_\mathrm{dc}=0.25$, $\alpha=0.7$) used in the remainder of this work.}
    \label{fig:static_phidc_alpha}
\end{figure}

We first consider the static limit, $\phi_\mathrm{ac}=0$, which serves as the dc reference for the ac-driven response. Figure~\ref{fig:static_phidc_alpha} shows the two-dimensional map  $\eta(\phi_\mathrm{dc},\alpha)$ for the conventional asymmetric SQUID with two purely sinusoidal CPRs. The map is antisymmetric under $\phi_\mathrm{dc}\to-\phi_\mathrm{dc}$, with $\eta$ vanishing at the time-reversal-invariant flux values $\phi_\mathrm{dc}=0,\pm 1/2$ and for identical junctions ($\alpha=1$), consistent with the symmetry analysis of Sec.~\ref{sec:analytics_summary}. The diode response is concentrated in narrow bands centered at half-integer flux quanta, with extrema reaching $|\eta|\approx 0.3$ for moderate asymmetry $\alpha\approx 0.7$. This localization into narrow bands deviates from the leading-order prediction of Eq.~(\ref{eq:eta_L}) and reflects the breakdown of the small-$\beta_\mathrm{L}$ expansion when the effective first harmonic $R_0$ becomes small. The four CPR cases of interest are already distinguishable in the static limit: (a) pure amplitude asymmetry, $i_\mathrm{c2}^{(1)}=i_\mathrm{c2}^{(2)}=0$; (b) subharmonic only, $i_\mathrm{c2}^{(1)}=0.5$ and $i_\mathrm{c2}^{(2)}=0$; (c) second harmonic only, $i_\mathrm{c2}^{(1)}=0$ and $i_\mathrm{c2}^{(2)}=0.5$; and (d) the mixed case, $i_\mathrm{c2}^{(1)}=i_\mathrm{c2}^{(2)}=0.5$.

\begin{figure}[t]
    \centering
    \includegraphics[width=1\linewidth]{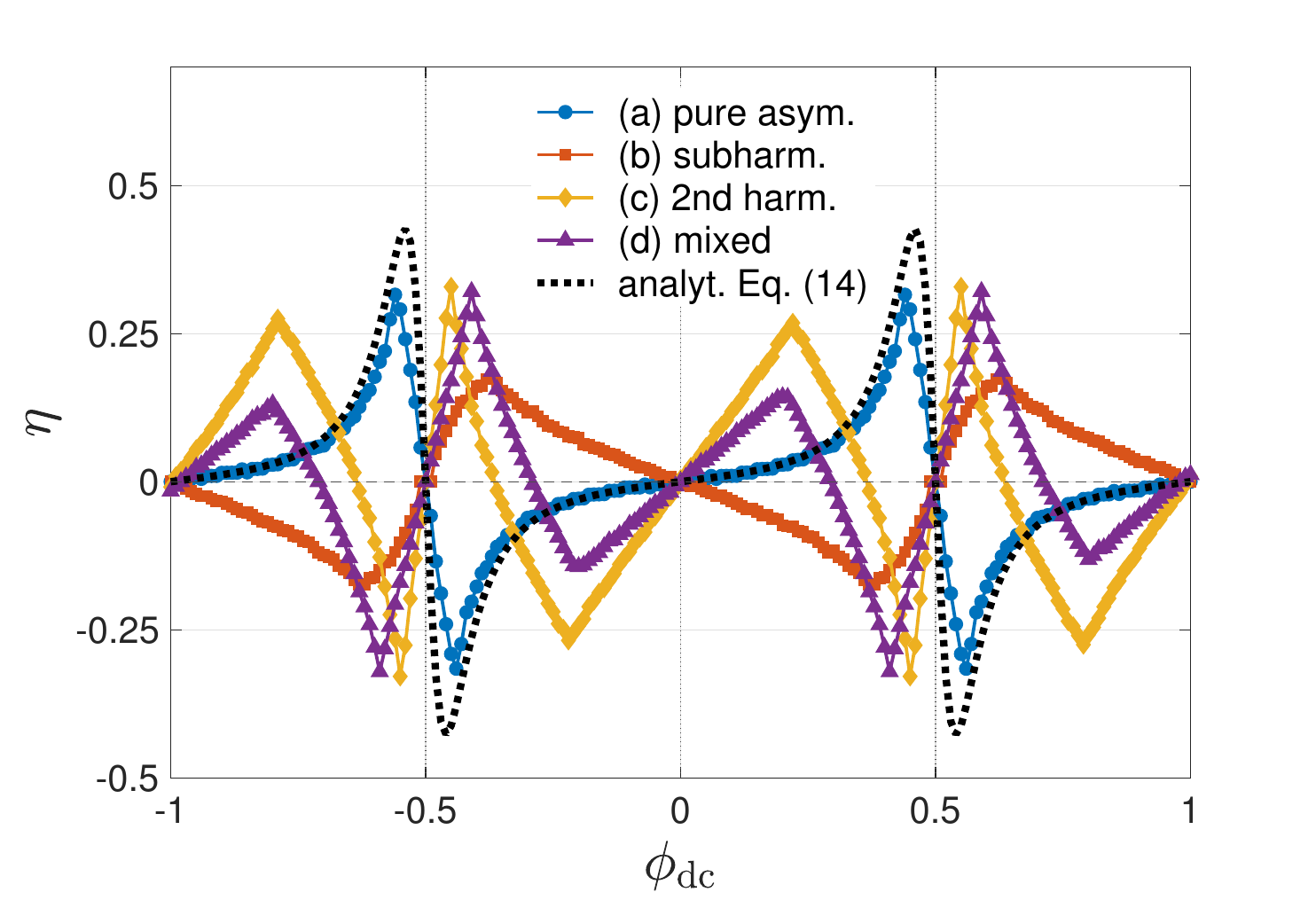}
    \caption{Static diode efficiency $\eta(\phi_\mathrm{dc})$ at     $\phi_\mathrm{ac}=0$, $\beta_\mathrm{c}=1$, $\beta_\mathrm{L}=0.15$, and     $\alpha=0.7$, for the four CPR cases defined by Eq.~(\ref{eq:CPR2}), as indicated in the legend. The dotted black curve shows the analytical prediction from Eq.~(\ref{eq:eta_L}). Vertical dotted lines mark the time-reversal symmetric flux values  $\phi_\mathrm{dc}=0$ and $\phi_\mathrm{dc}=\pm 1/2$.}
\label{fig:static_phidc_lines}
\end{figure}

Figure~\ref{fig:static_phidc_lines} shows $\eta(\phi_\mathrm{dc})$ at $\alpha=0.7$ for these four CPR cases (a)-(d). All four curves are antisymmetric under $\phi_\mathrm{dc}\to-\phi_\mathrm{dc}$ and vanish at $\phi_\mathrm{dc}=0,\pm 1/2,\pm 1$, consistent with the symmetry analysis. The four cases exhibit qualitatively distinct static fingerprints. Case~(a) exhibits a single sharp peak per half-period near $\phi_\mathrm{dc}\approx\pm 0.55$, where $R_0$ is suppressed. The analytical prediction, Eq.~(\ref{eq:eta_L}) (dotted black curve), captures the peak position and overall shape but overestimates its amplitude, reflecting the breakdown of the small-$\beta_\mathrm{L}$ expansion discussed in Sec.~\ref{sec:analytics_summary}. Case~(b) displays a broader, smoother response with extrema near $\phi_\mathrm{dc}\approx\pm 0.4$, characteristic of the $4\pi$-periodic $\sin(\varphi/2)$ component. Case~(c) develops two extrema per half-period---one maximum and one minimum---compared with the single extremum in case~(a), reflecting the $\pi$-periodic structure of $\sin 2\varphi$ and the resulting doubling of features in the static response. Case~(d) retains the two-extremum structure of case~(c), with the amplitudes and positions of the extrema modified by the subharmonic contribution of case~(b). 

\begin{figure}[t]
    \centering
    \includegraphics[width=1\linewidth]{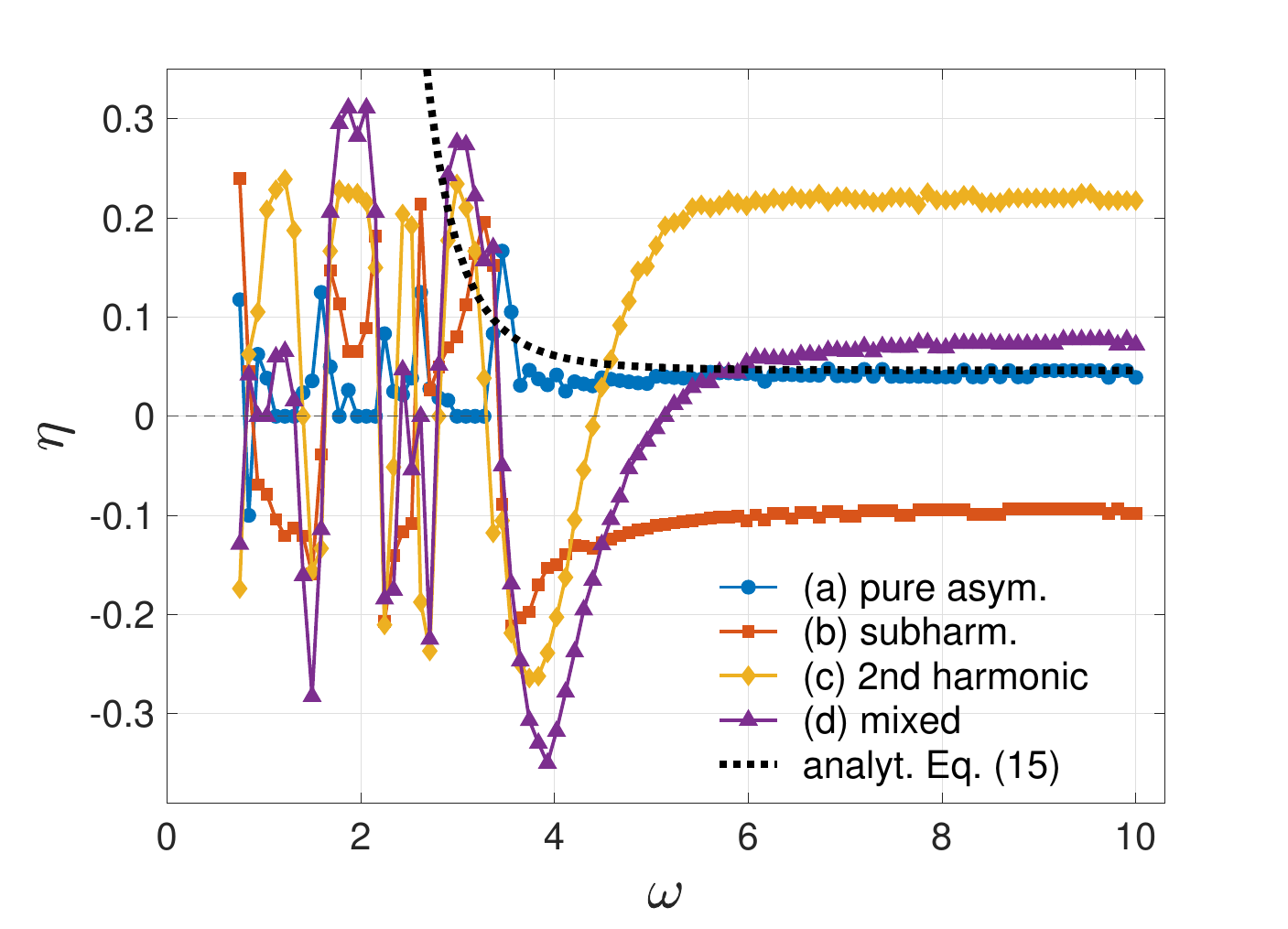}
    \caption{Frequency dependence of the diode efficiency
    $\eta(\omega)$ at fixed $\phi_\mathrm{ac}=1.5$, $\phi_\mathrm{dc}=0.25$,
    $\beta_\mathrm{c}=1$, $\beta_\mathrm{L}=0.15$, and $\alpha=0.7$, for the four CPR cases, as indicated. The dotted black curve shows the analytical prediction from Eq.~(\ref{eq:eta_ac}).}
    \label{fig:eta_vs_omega}
\end{figure}

The high-frequency behavior of the system is summarized in Fig.~\ref{fig:eta_vs_omega}, which shows $\eta(\omega)$ at fixed $\phi_\mathrm{ac}=1.5$ for all four CPR cases. Three distinct frequency regimes can thus be identified in the dependence $\eta(\omega)$. Namely, for $0.75<\omega<3$, the response is strongly nonperturbative, exhibiting aperiodic oscillations of $\eta$ associated with multiple crossings of Bessel-function zeros along the working line $\phi_\mathrm{ac}=1.5$. In the transition region $3<\omega<5$, the oscillations are progressively damped as the system approaches the adiabatic limit. For $\omega \gtrsim 5$, each curve saturates to a distinct plateau: case~(a) approaches $\eta \approx 0.05$, close to the static value $\eta_\mathrm{L} \approx 0.046$ from Eq.~(\ref{eq:eta_L}); case~(b) saturates at $\eta \approx -0.1$; case~(c) at $\eta \approx 0.22$; and case~(d) at $\eta \approx 0.08$. These asymptotic values provide a compact CPR fingerprint: a single measurement at sufficiently large $\omega$ is sufficient to unambiguously distinguish all four cases. The dotted black curve in Fig.~\ref{fig:eta_vs_omega} shows the analytical prediction of Eq.~(\ref{eq:eta_ac}) for case~(a). It diverges near $\omega \approx \omega_\mathrm{L}$, where the perturbative expansion in $\xi_0$ breaks down, but recovers the correct adiabatic plateau for $\omega \gtrsim 5$, in quantitative agreement with the numerical results for case~(a).

\begin{figure*}[t]
    \centering
    \includegraphics[width=0.5\linewidth]{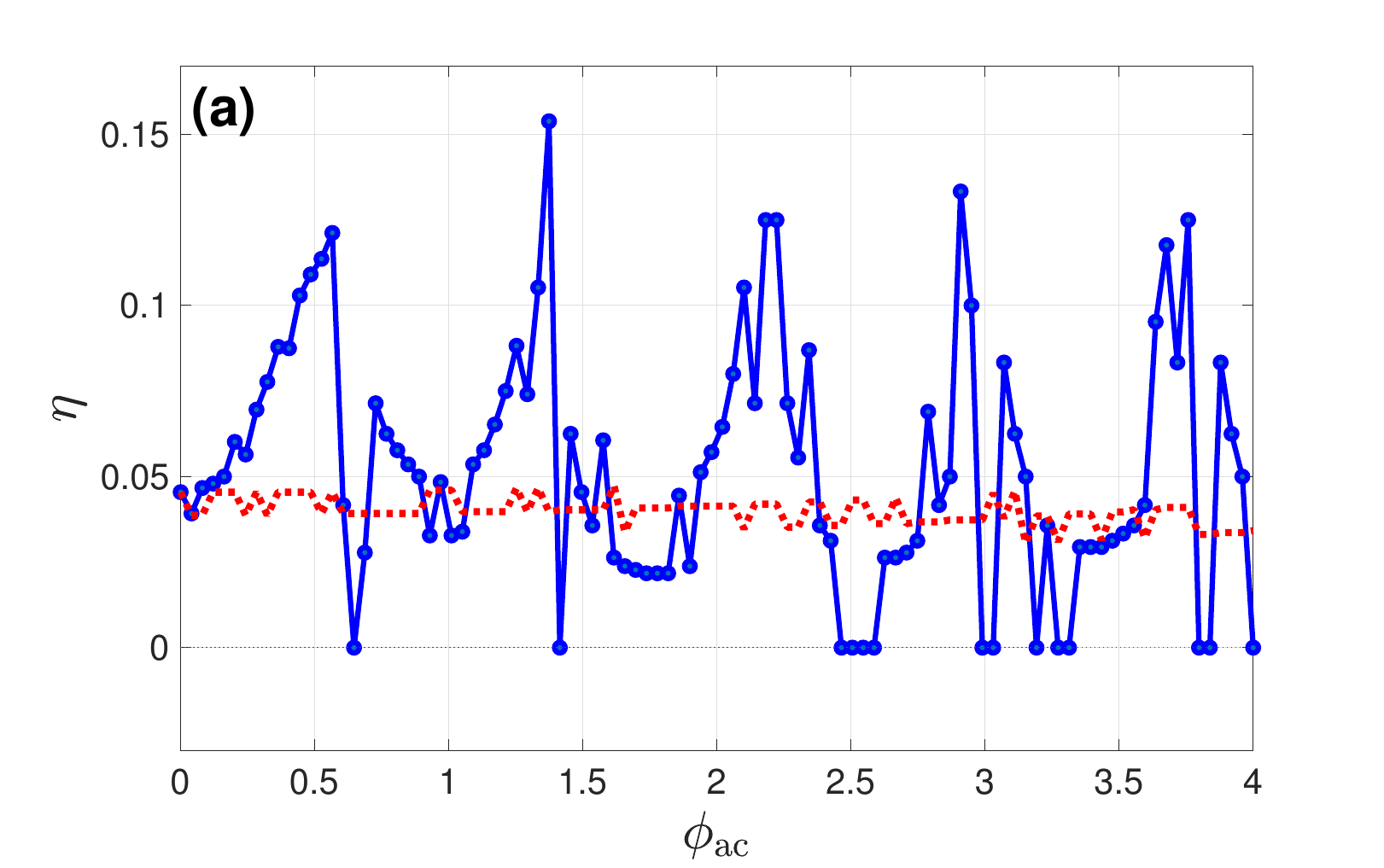}\hfil
    \includegraphics[width=0.5\linewidth]{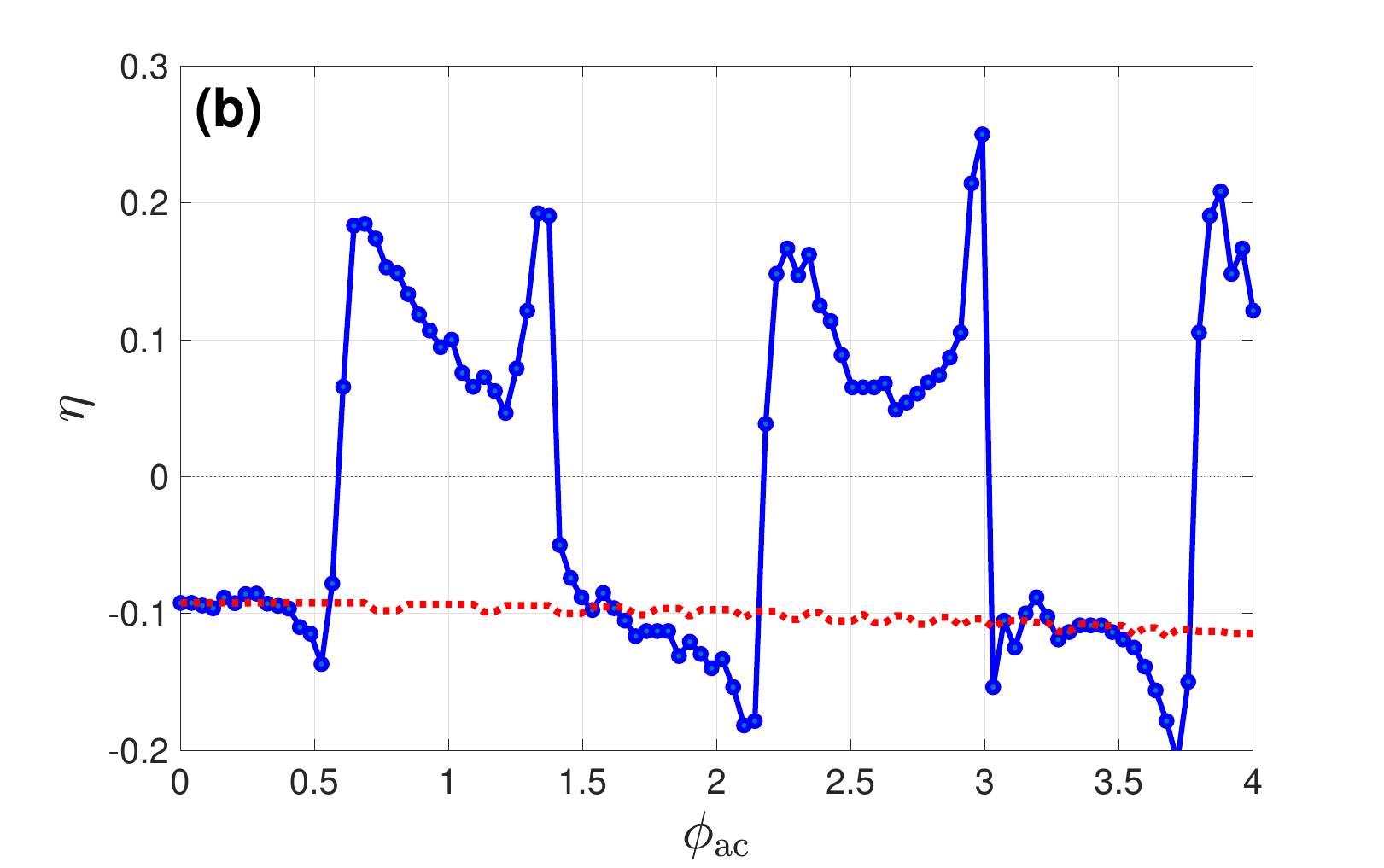}\par\medskip
    \includegraphics[width=0.5\linewidth]{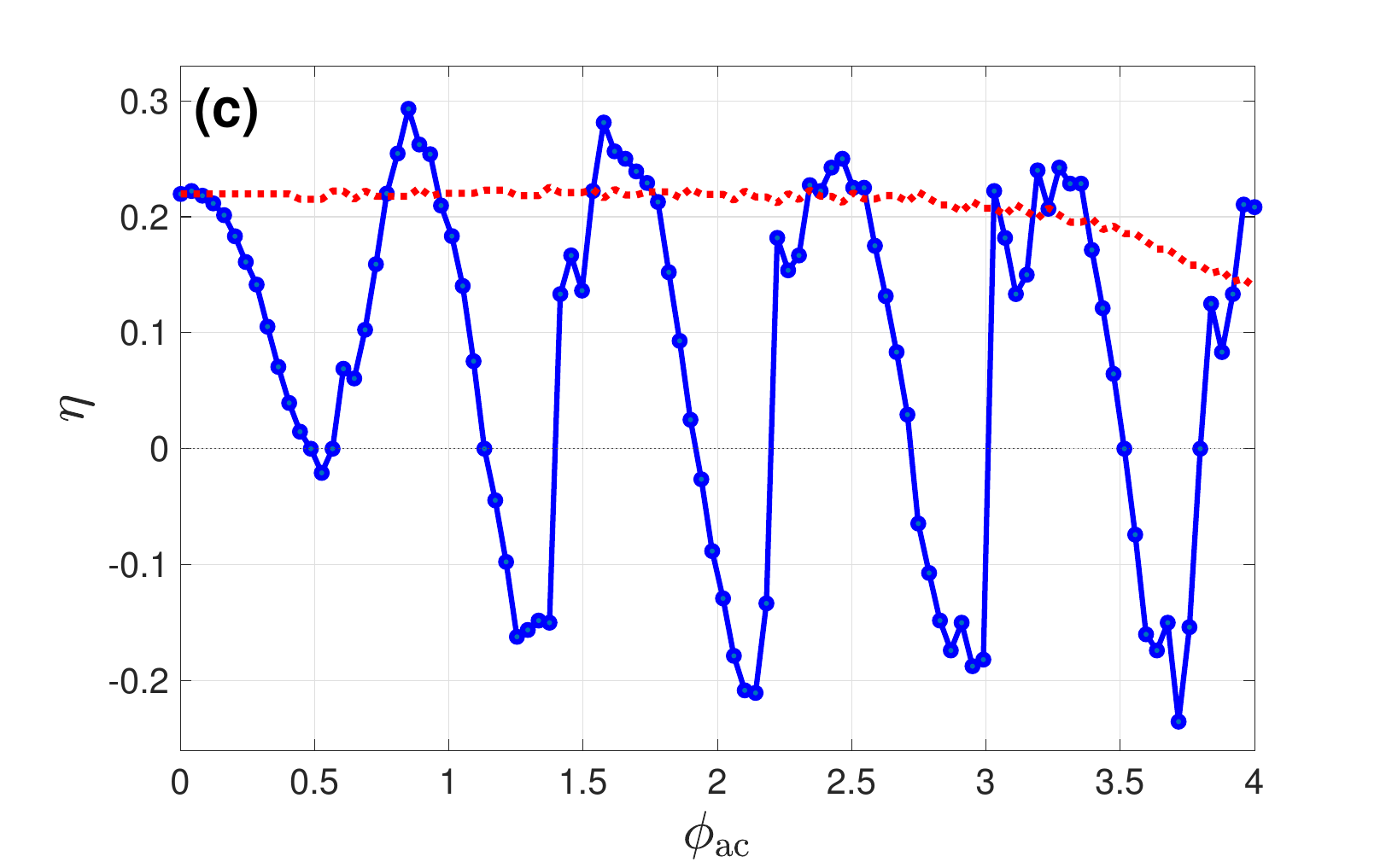}\hfil
    \includegraphics[width=0.5\linewidth]{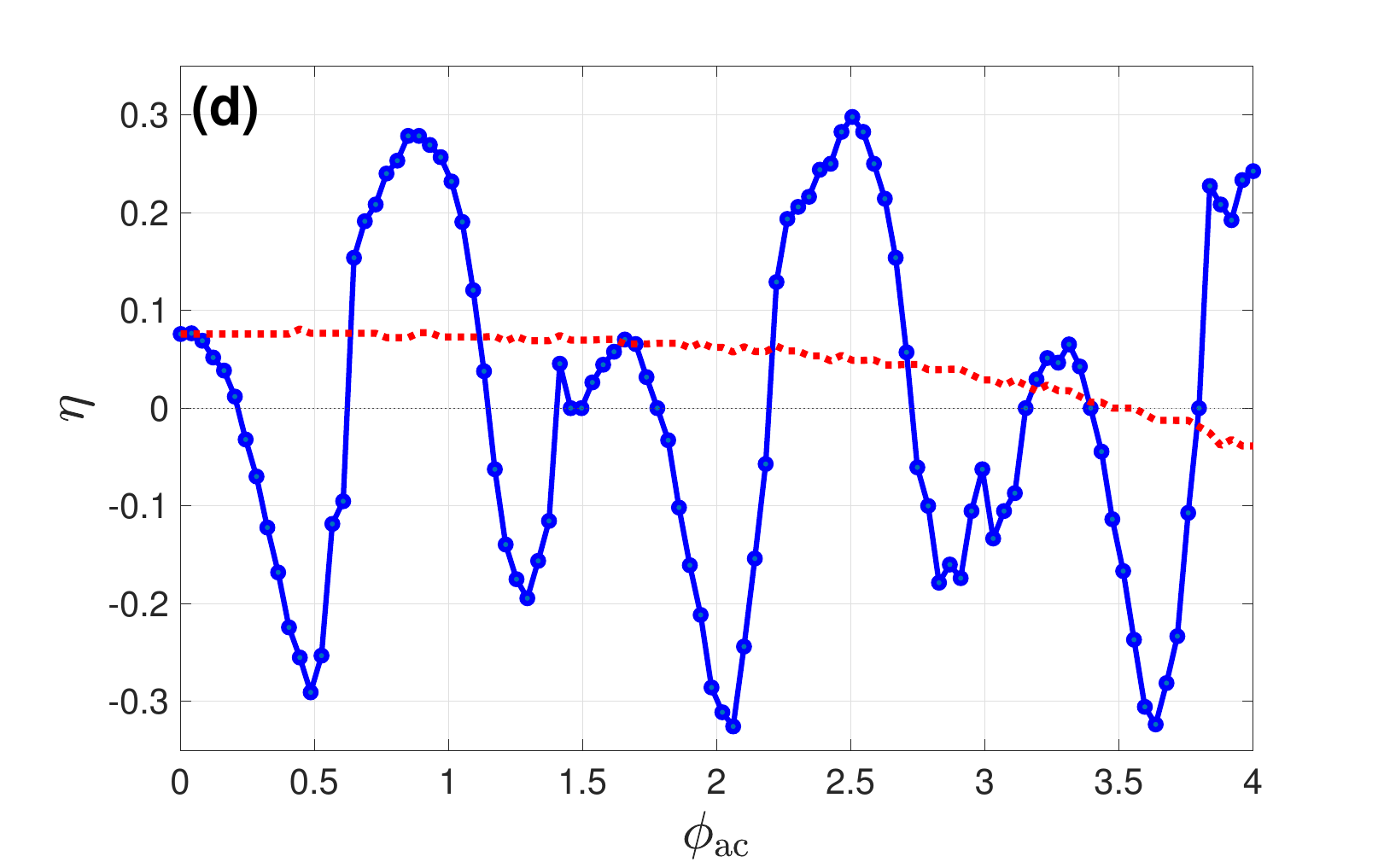}
    \caption{Flux-drive-amplitude dependence of the diode efficiency $\eta(\phi_\mathrm{ac})$ at $\phi_\mathrm{dc}=0.25$, $\beta_\mathrm{c}=1$, $\beta_\mathrm{L}=0.15$, and $\alpha=0.7$ for four CPR cases: (a) pure amplitude asymmetry, (b) subharmonic only, (c) second-harmonic only, and (d) mixed harmonics. Each panel shows two curves corresponding to $\omega=1$ (solid blue with markers) and the adiabatic limit $\omega=7.5$ (dotted red).}
\label{fig:eta_vs_phiac}
\end{figure*}

The complementary dependence $\eta(\phi_\mathrm{ac})$ at fixed $\omega$ provides a particularly direct experimental implementation of the fingerprinting protocol. Figure~\ref{fig:eta_vs_phiac} shows $\eta(\phi_\mathrm{ac})$ for the four CPR cases at two representative frequencies: a subresonant value $\omega=1$ (well below $\omega_\mathrm{L}\approx 2.06$) and an adiabatic value $\omega=7.5$ (well above $\omega_\mathrm{L}$). The contrast between the two frequencies directly reflects the analytical predictions of Eq.~(\ref{eq:eta_jacobi_anger_main}).
The four cases produce qualitatively distinct patterns: case~(a) shows essentially no polarity reversals, consistent with the absence of higher CPR harmonics; case~(b) exhibits approximately five sign changes governed by the zeros of $J_0(\xi_0/2)$; case~(c) shows approximately seven sign changes governed by the zeros of $J_0(2\xi_0)$; and case~(d) displays the densest structure, with up to a dozen sign changes and a characteristic beat-like envelope arising from interference between subharmonic and second-harmonic contributions in Eq.~(\ref{eq:eta_jacobi_anger_main}).
For $\omega=7.5$, the Bessel arguments remain small, and the curves closely follow the static baselines in Fig.~\ref{fig:eta_vs_phiac}, with only case~(d) exhibiting a sign change within the plotted range.

\section{AC-driven phase diagrams and CPR fingerprints}
\label{sec:numerics}
\subsection{Parameters choice and justification}

We now turn to numerical solutions of the full coupled equations of motion, Eqs.~(\ref{eq:EOM1})--(\ref{eq:coupling}), for the four representative CPR scenarios of Eq.~(\ref{eq:CPR2}): (a) pure amplitude asymmetry, $i_\mathrm{c2}^{(1)}=i_\mathrm{c2}^{(2)}=0$; (b) subharmonic only, $i_\mathrm{c2}^{(1)}=0.5$ and $i_\mathrm{c2}^{(2)}=0$; (c) second harmonic only, $i_\mathrm{c2}^{(1)}=0$ and $i_\mathrm{c2}^{(2)}=0.5$; and (d) the mixed case, $i_\mathrm{c2}^{(1)}=i_\mathrm{c2}^{(2)}=0.5$.
Throughout this section we fix $\beta_\mathrm{c}=1$, $\beta_\mathrm{L}=0.15$, $\phi_\mathrm{dc}=0.25$, and $\alpha=i_\mathrm{c2}^{(0)}=0.7$. These values are representative rather than fine-tuned. The amplitude asymmetry $\alpha=0.7$ places the SQUID in an intermediate-asymmetry regime. It is sufficiently strong to generate a robust diode response, while it is far enough from the symmetry-protected line $\alpha=1$ to avoid suppression of $\eta$.

The exotic-CPR amplitudes $i_\mathrm{c2}^{(1,2)}=0.5$ are also in an intermediate regime. They are small enough that the leading-order dynamics remain dominated by the $\sin\varphi$ harmonic, yet large enough to produce clear signatures. The choice $\phi_\mathrm{dc}=0.25$ corresponds to the maximum-slope point of the static $\eta(\phi_\mathrm{dc})$ curve in Fig.~\ref{fig:static_phidc_lines}, away from the half-integer flux quanta where the perturbative description breaks down. The ac frequency axis spans $\omega\in[0.75,5]$, which includes the inductive resonance $\omega_\mathrm{L}=\sqrt{2/(\pi\beta_\mathrm{c}\beta_\mathrm{L})}\approx 2.06$ for our parameters as well as the adiabatic high-frequency regime, while the ac amplitude axis spans $\phi_\mathrm{ac}\in[0.01,3]$. We do not extend the frequency range below $\omega=0.75$, since in that regime the ac period becomes comparable to or longer than the relaxation time following a current-bias step, and the extracted critical current at fixed $(\phi_\mathrm{ac},\omega)$ no longer admits a clean stroboscopic interpretation. The quasi-static limit $\omega\to 0$ is instead described by the static finite-inductance result of Eq.~(\ref{eq:eta_L}) and shown in the static maps of Figs.~\ref{fig:static_phidc_alpha} and~\ref{fig:static_phidc_lines}. The numerical procedure used to extract $\eta$ for each $(\phi_\mathrm{ac},\omega)$ point is described in~\cite{SupplementalMaterial}.

\begin{figure*}[t]
\centering
    \includegraphics[width=0.35\linewidth]{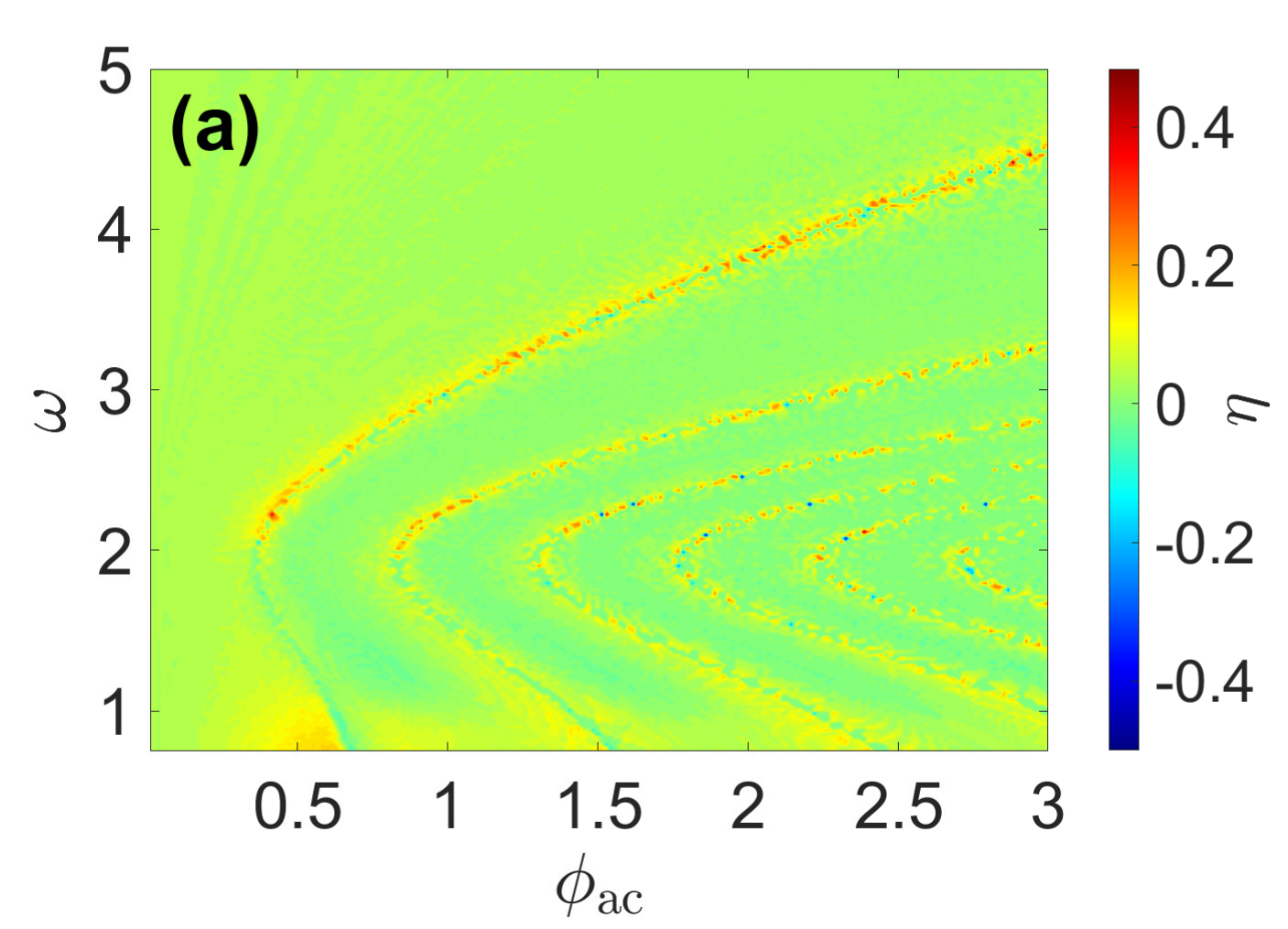}\hfil
    \includegraphics[width=0.35\linewidth]{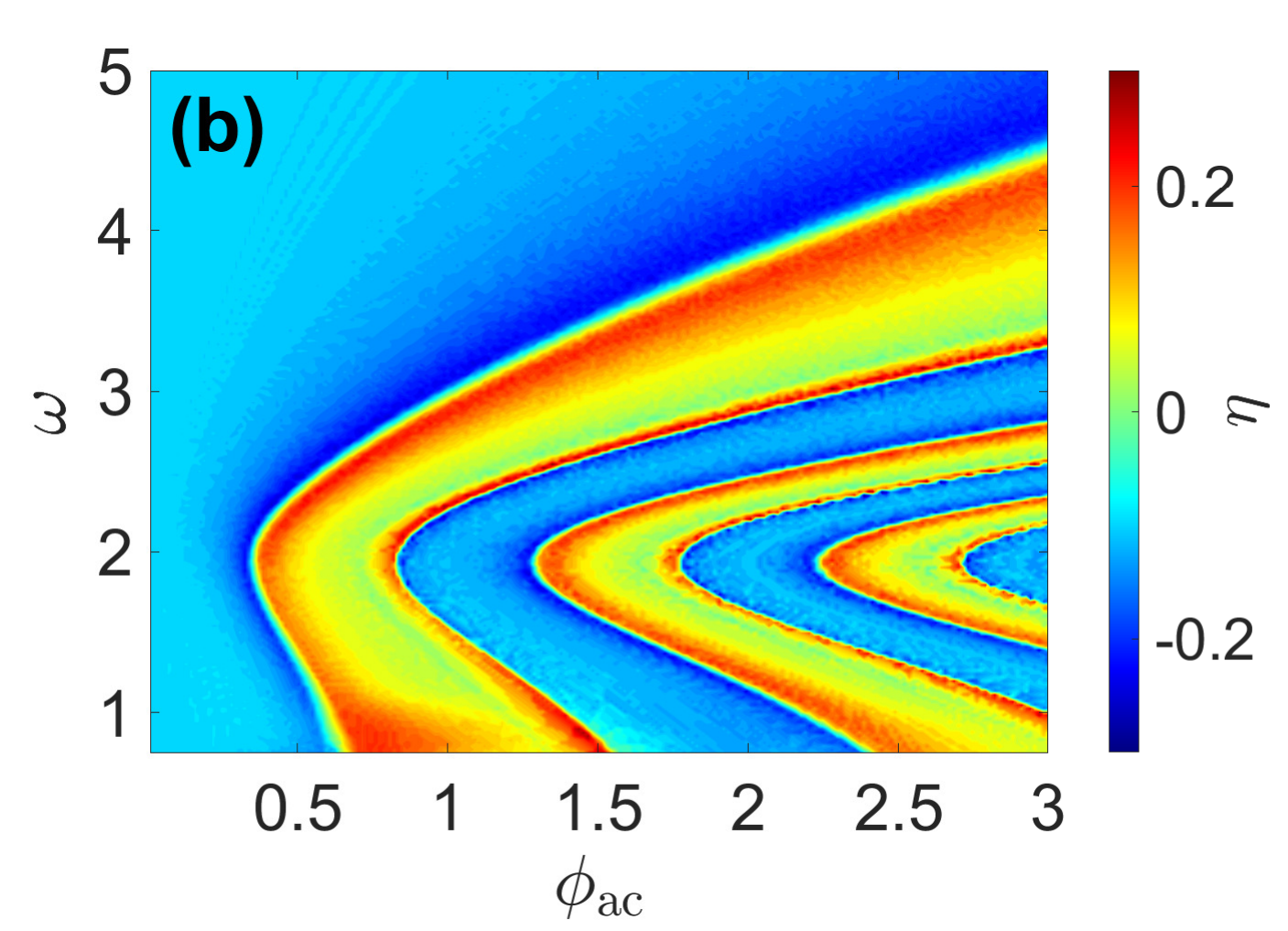}\par\medskip
    \includegraphics[width=0.35\linewidth]
{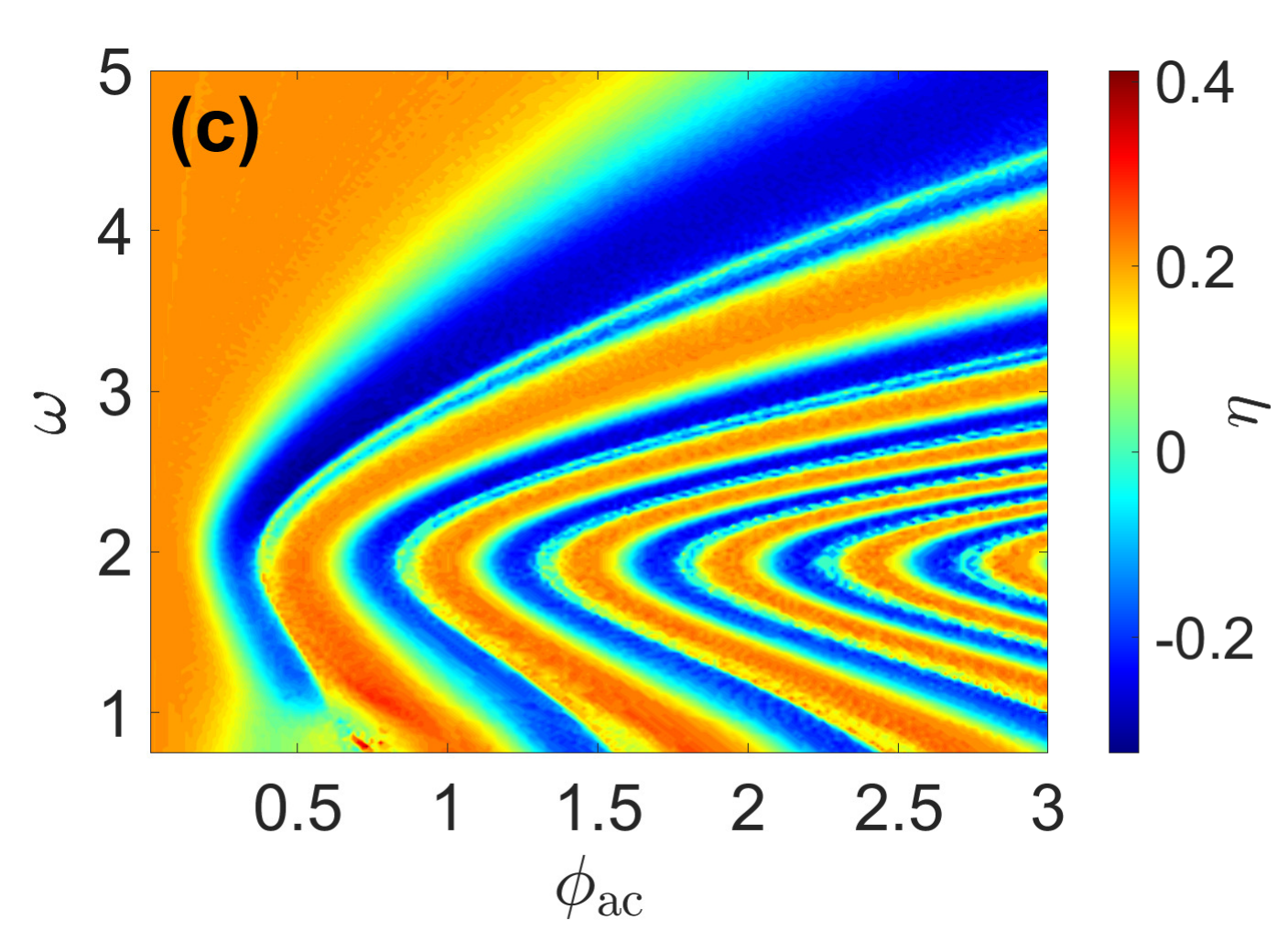}\hfil
    \includegraphics[width=0.35\linewidth]
{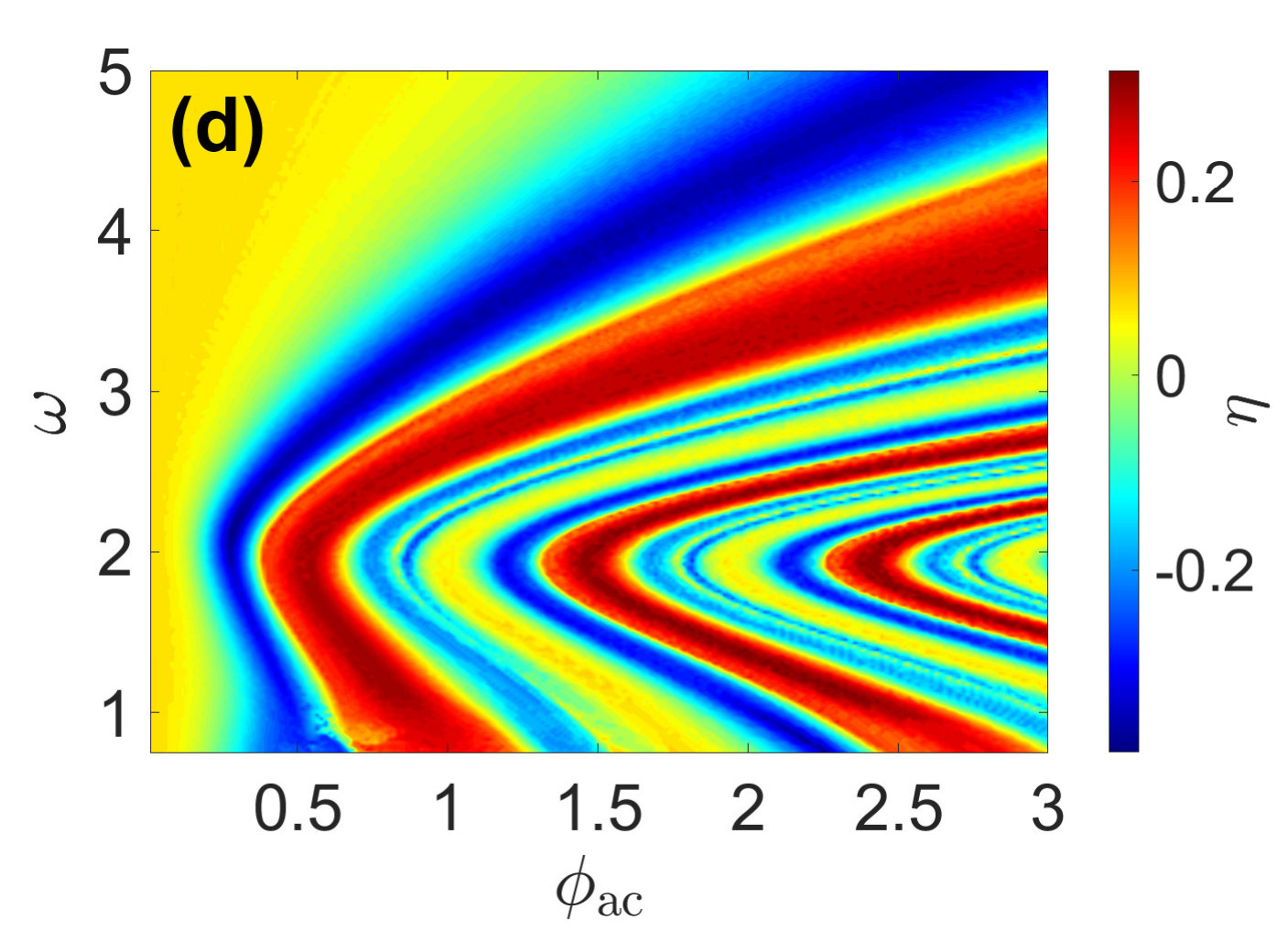}
    \caption{Two-dimensional phase diagrams of the diode efficiency $\eta(\phi_\mathrm{ac},\omega)$ for a dc SQUID with $\beta_\mathrm{c}=1.0$, $\beta_\mathrm{L}=0.15$, $\phi_\mathrm{dc}=0.25$,
    $i_\mathrm{c2}^{(0)}=0.7$, and (a)~$i_\mathrm{c2}^{(1)}=i_\mathrm{c2}^{(2)}=0$, (b)~$i_\mathrm{c2}^{(1)}=0.5$, $i_\mathrm{c2}^{(2)}=0$, (c)~$i_\mathrm{c2}^{(1)}=0$, $i_\mathrm{c2}^{(2)}=0.5$, (d)~$i_\mathrm{c2}^{(1)}=i_\mathrm{c2}^{(2)}=0.5$.}
    \label{fig:phase_diagrams}
\end{figure*}

\subsection{Pure amplitude asymmetry}
\label{sec:diag_a}

Figure~\ref{fig:phase_diagrams}(a) shows the $\eta(\phi_\mathrm{ac},\omega)$ phase diagram for the conventional asymmetric SQUID with purely sinusoidal CPRs ($i_\mathrm{c2}^{(1)}=i_\mathrm{c2}^{(2)}=0$). Compared with the exotic-CPR cases discussed below, the response is weak across the entire $(\phi_\mathrm{ac},\omega)$ plane, with $|\eta|$ remaining substantially smaller than in panels~(b)-(d) over most of the map. The strongest response follows a locus near $\omega\approx\omega_\mathrm{L}\approx 2.06$ of approximately constant fast-oscillation amplitude $\xi_0$, consistent with the resonant enhancement of the ac-induced contribution $\Delta\eta_\mathrm{ac}$ predicted by Eq.~(\ref{eq:eta_ac}). The map is not strictly featureless: a faint family of arc-like features is resolved along loci of constant $\xi_0$, in agreement with the non-perturbative expression Eq.~(\ref{eq:eta_a_nonpert}). Although these features are parametrically weaker than the arcs observed for the exotic-CPR cases, their overall scale is set by the small factor $K_{\mathrm{tot}}$. They also lie below the resolution of coarser numerical grids, which is why the leading-order analysis treats case~(a) as the reference response. 

The key point for the fingerprinting protocol is that the case-(a) arcs are governed by $J_0(\xi_0)$ rather than by $J_0(\xi_0/2)$ or $J_0(2\xi_0)$. Consequently, their spacing differs characteristically from that of the subharmonic and second-harmonic cases, while their amplitude remains strongly suppressed. This map therefore serves as the reference response against which the enhanced amplitude and modified arc spacing produced by exotic CPR components can be identified.

\subsection{Subharmonic CPR}
\label{sec:diag_b}

Introducing a $4\pi$-periodic subharmonic $\sin(\varphi/2)$
contribution in $J_2$ significantly modifies the phase
diagram, as shown in Fig.~\ref{fig:phase_diagrams}(b) for
$i_\mathrm{c2}^{(1)}=0.5$ and $i_\mathrm{c2}^{(2)}=0$. A family of
widely spaced concentric arc-shaped bands now fills the
$(\phi_\mathrm{ac},\omega)$ plane, with adjacent bands having
opposite diode polarity. 
The wide arc spacing is the hallmark of the half-argument Bessel function $J_0(\xi_0/2)$ predicted by Eq.~(\ref{eq:eta_jacobi_anger_main}). Its zeros occur at
larger values of $\xi_0$ (and hence of $\phi_\mathrm{ac}$) than
those of the corresponding $J_0(\xi_0)$, directly
reflecting the $4\pi$-periodicity of the subharmonic CPR.
The arcs curve upward with increasing $\omega$ at low
frequencies and flatten above $\omega\approx 3.5$ as the
system approaches the adiabatic high-frequency limit.

\subsection{Second-harmonic CPR}
\label{sec:diag_c}

A qualitatively similar but quantitatively distinct
pattern emerges when the subharmonic contribution is
replaced by a $\pi$-periodic second harmonic $\sin 2\varphi$
in $J_2$, as shown in Fig.~\ref{fig:phase_diagrams}(c) for
$i_\mathrm{c2}^{(1)}=0$ and $i_\mathrm{c2}^{(2)}=0.5$. The arc structure
is now denser. 
This increased density reflects the
doubled Bessel function argument $J_0(2\xi_0)$ predicted
by Eq.~(\ref{eq:eta_jacobi_anger_main}). Its zeros are
spaced at half the $\xi_0$ distance compared to
$J_0(\xi_0/2)$, consistent with the $\pi$ periodicity of
$\sin 2\varphi$ being half that of $\sin(\varphi/2)$. 
The arc boundaries show  more irregularity than in panel~(b), which we attribute to the faster phase oscillations associated with $\sin 2\varphi$ approaching the limits of the numerical
resolution.

\subsection{Mixed CPR}
\label{sec:diag_d}

The most intriguing behavior appears when both subharmonic
and second harmonic contributions are present
simultaneously, as shown in Fig.~\ref{fig:phase_diagrams}(d) for $i_\mathrm{c2}^{(1)}=i_\mathrm{c2}^{(2)}=0.5$. The phase diagram
exhibits a notable feature: the
mixed case shows fewer visible arcs than panel~(c)
alone, rather than the larger number one might naively
expect from a simple superposition. 
This nonlinear intermodulation is anticipated by
Eq.~(\ref{eq:eta_jacobi_anger_main}): when both terms in
the formula contribute, the corresponding $J_0(\xi_0/2)$
and $J_0(2\xi_0)$ Bessel factors can interfere
constructively or destructively at different
$(\phi_\mathrm{ac},\omega)$ points, partially canceling each
other's polarity-switching boundaries and reducing the
total arc count. This genuine nonlinear effect demonstrates that the $\eta(\phi_\mathrm{ac},\omega)$ map encodes not only the presence of individual CPR components, but also their mutual interference within the coupled nonlinear dynamics.

\subsection{Overdamped and underdamped regimes}
\label{sec:diag_betac}

\begin{figure*}[t]
    \centering
    \includegraphics[width=0.77\linewidth]{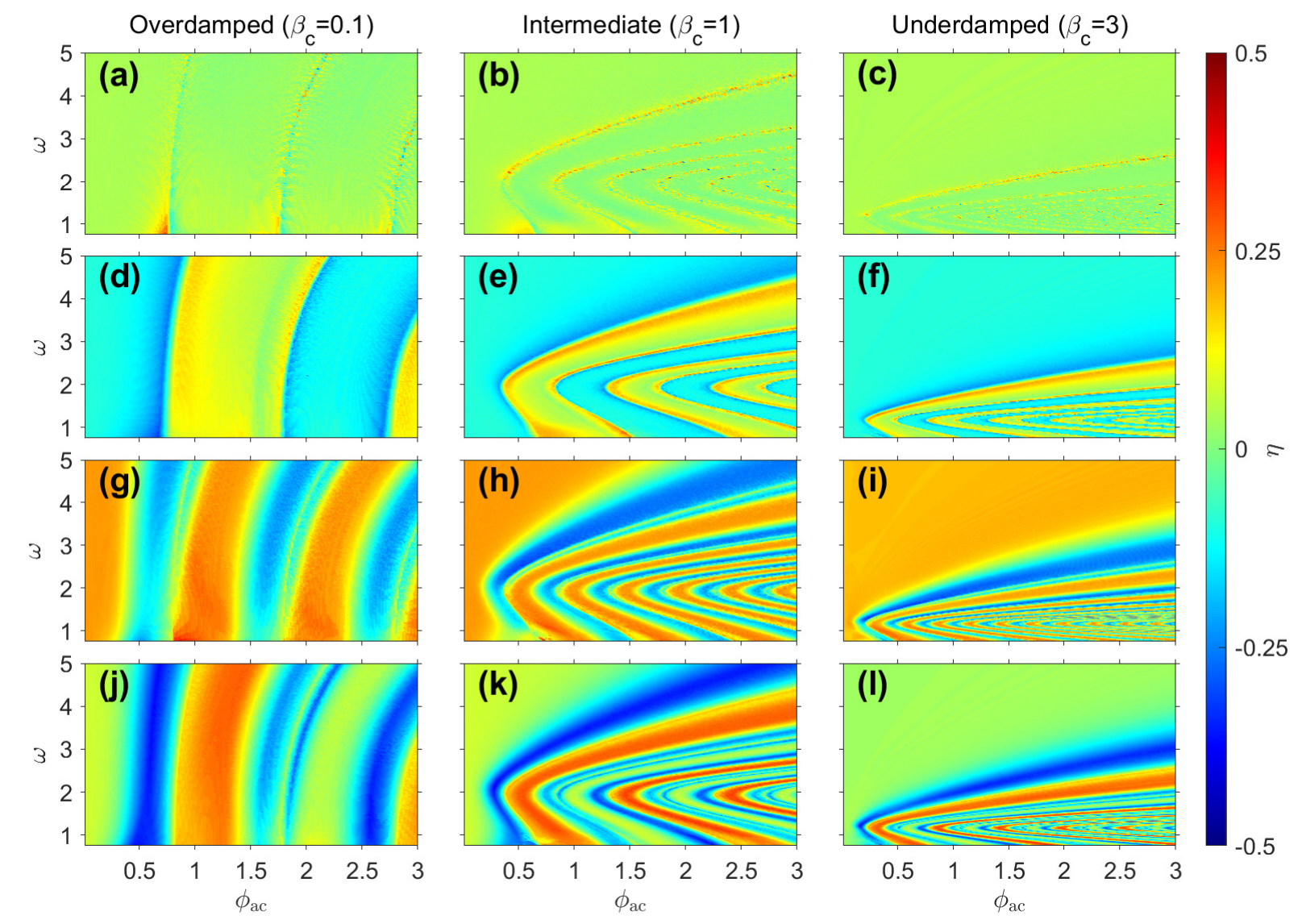}
    \caption{Two-dimensional phase diagrams of the diode efficiency $\eta(\phi_\mathrm{ac},\omega)$ for a dc SQUID across the overdamped-to-underdamped crossover. Columns correspond to the overdamped ($\beta_\mathrm{c}=0.1$, left), intermediate ($\beta_\mathrm{c}=1$, center), and underdamped ($\beta_\mathrm{c}=3$, right) regimes. Rows correspond to the four CPR cases of Eq.~(\ref{eq:CPR2}): (a)-(c) pure amplitude asymmetry, (d)-(f) subharmonic only, (g)-(i) second harmonic only, and (j)-(l) the mixed case. Other parameters are as in Fig.~\ref{fig:phase_diagrams}: $\phi_\mathrm{dc}=0.25$, $\alpha=0.7$, and $\beta_\mathrm{L}=0.15$.}
    \label{fig:phase_matrix}
\end{figure*}

The phase diagrams of Fig.~\ref{fig:phase_diagrams} were
computed at the intermediate value $\beta_\mathrm{c}=1$. Since the
Stewart--McCumber parameter controls both the damping of the
difference mode ($\gamma=1/\beta_\mathrm{c}$) and the position of the
inductive resonance [$\omega_\mathrm{L}=\sqrt{2/(\pi\beta_\mathrm{c}\beta_\mathrm{L})}$], it is important to check that the fingerprinting distinctions persist away from this reference point. We therefore performed the full $2\times4$ set of calculations for an overdamped regime, $\beta_\mathrm{c}=0.1$, and an underdamped regime, $\beta_\mathrm{c}=3$, and compare them with the intermediate case in Fig.~\ref{fig:phase_matrix}. The underdamped case corresponds to a moderately underdamped junction with quality factor $Q=\omega_\mathrm{L}\beta_\mathrm{c}=\sqrt{2\beta_\mathrm{c}/(\pi\beta_\mathrm{L})}\approx 3.6$. We do not consider the strongly hysteretic regime $\beta_\mathrm{c}\gg 1$, where switching-current extraction becomes history dependent and the stroboscopic protocol employed here no longer applies straightforwardly.

In the overdamped column [Fig.~\ref{fig:phase_matrix}(a),(d),(g),(j)], the resonance $\omega_\mathrm{L}\approx 6.5$ lies above the scanned frequency range, rendering $\xi_0$ nearly frequency independent. As a result, the arcs are nearly vertical, and the frequency axis contributes little information beyond that contained in the drive-amplitude dependence. The four CPR cases nevertheless remain readily distinguishable by the density and amplitude of their arcs: case~(a) exhibits weak, closely spaced features on a low-amplitude background; case~(b) shows a few widely spaced, high-contrast bands; case~(c) displays a much denser band structure; and case~(d) combines an intermediate density with the irregular pattern characteristic of intermodulation.

In the underdamped column [Fig.~\ref{fig:phase_matrix}(c),(f),(i),(l)], the resonance $\omega_\mathrm{L}\approx 1.19$ lies within the scanned window. The arcs therefore fan out from a focal region near $(\phi_\mathrm{ac}\to 0,,\omega\approx\omega_\mathrm{L})$ and become increasingly dense at small $\phi_\mathrm{ac}$, as the resonant enhancement compresses multiple Bessel zeros into a narrow drive-amplitude range. At high frequencies the arcs disappear and the maps become smooth. The ordering of arc densities across the four CPR cases is preserved, with the contrast between the sparse case-(b) pattern and the dense case-(c) pattern even more pronounced than at $\beta_\mathrm{c}=1$. The weak case-(a) arcs remain visible [Fig.~\ref{fig:phase_matrix}(c)], now curved and concentrated near the resonance, consistent with the $J_0(\xi_0)$ dependence of Eq.~(\ref{eq:eta_a_nonpert}). Taken together, the results for $\beta_\mathrm{c}=0.1$, $1$, and $3$ demonstrate that the fingerprinting signatures are robust across the overdamped-to-underdamped crossover, with the frequency dimension being most informative when the resonance lies within the accessible frequency window.
\section{Discussion}
\subsection{Main results}
\label{sec:diag_summary}
 
Taken together, the phase diagrams of the diode efficiency in Fig.~\ref{fig:phase_diagrams}, the frequency cuts in Fig.~\ref{fig:eta_vs_omega}, and the drive-amplitude cuts in Fig.~\ref{fig:eta_vs_phiac} provide a practical route to distinguishing several classes of CPRs. Namely, a weak, strongly suppressed response with arcs governed by $J_0(\xi_0)$ identifies a conventional amplitude-asymmetric dc SQUID, while widely spaced high-amplitude arcs emerging at small $\phi_\mathrm{ac}$ signal a $4\pi$-periodic subharmonic component. Denser arc patterns indicate a $\pi$-periodic second-harmonic contribution, whereas intermodulation features---fewer arcs in the 2D maps but the highest density of polarity switches in 1D cuts---reveal the coexistence of multiple CPR harmonics.

The advantage of this fingerprinting protocol rests on the
analytical structure uncovered in Sec.~\ref{sec:analytics_summary}. The Jacobi-Anger reduction
of the fast-driven dynamics shows that the ac modulation
enters the effective CPR through three distinct
Bessel-function dressings: $J_0(\xi_0)$ for the
conventional $\sin\varphi$ harmonic, $J_0(\xi_0/2)$ for the
$4\pi$-periodic subharmonic, and $J_0(2\xi_0)$ for the
$\pi$-periodic second harmonic. Each dressing carries the
microscopic periodicity of its underlying CPR component
directly into the $(\phi_\mathrm{ac},\omega)$ plane. The arc-like nodal lines that populate the numerical phase diagrams are nothing other
than the loci of these Bessel-function zeros under the
mapping $(\phi_\mathrm{ac},\omega)\mapsto\xi_0(\phi_\mathrm{ac},\omega)$
defined by Eq.~(\ref{eq:xi0_main}). Thus, the interval between Bessel zeros serves as a transport-level indicator reflecting
the microscopic periodicity of CPR. Note,  similar reasoning has previously been used to explain arc structures in the resistance associated with the motion of a dc+ac driven Abrikosov vortex in a cosine pinning potential~\cite{Shk11prb}. This correspondence follows from the similarity between the resistively shunted Josephson junction model and the Langevin equation describing vortex motion, where the vortex coordinate maps onto the Josephson phase and the ac current drive maps onto the ac flux drive used here.
 
It is important to clarify the applicability limits of the present framework. The Jacobi--Anger reduction assumes $\beta_\mathrm{L}\ll 1$, such that the phase difference is inductively locked to the applied flux. For weakly screened loops ($\beta_\mathrm{L}\gtrsim 1$), the phase acquires its own dynamics and the simple Bessel dressing of Eq.~(\ref{eq:bessel_dressing_main}) is expected to be modified by additional mode-coupling effects. By contrast, the dependence on the Stewart--McCumber parameter $\beta_\mathrm{c}$ is not essential for the fingerprinting protocol. As shown in Sec.~\ref{sec:diag_betac}, the distinguishing features persist from the overdamped regime ($\beta_\mathrm{c}=0.1$) to the moderately underdamped regime ($\beta_\mathrm{c}=3$), with $\beta_\mathrm{c}$ mainly affecting the arc curvature through the position of the resonance $\omega_\mathrm{L}$. We do not address the strongly hysteretic regime $\beta_\mathrm{c}\gg 1$, where switching currents become history dependent and a sweep-direction-resolved protocol would be required.
Finally, near the nodal lines of $J_0(\xi_0)$ or in regimes where $R_0\to 0$, the perturbative expansion in the subharmonic and second-harmonic amplitudes breaks down, and the full nonlinear CPR must be treated without expansion. These regimes account for the quantitative deviations between numerical results and analytical predictions discussed in Secs.~\ref{sec:analytics_summary} and~\ref{sec:numerics}.

A further point concerns the scope of the  spectroscopy protocol. The method is designed to identify which CPR harmonics are present via the spacing and amplitude of the arcs, rather than to reconstruct the precise harmonic amplitudes $i_\mathrm{c2}^{(1,2)}$. In realistic junctions these amplitudes are unknown and generally unequal, so the mixed case-(d) pattern should be regarded as representative of a continuous family of responses. A quantitative reconstruction of the CPR would require fitting the full $\eta(\phi_\mathrm{ac},\omega)$ map to Eq.~(\ref{eq:eta_jacobi_anger_main}), with $\beta_\mathrm{c}$, $\beta_\mathrm{L}$, and $\omega_\mathrm{L}$ independently determined.

A further limitation concerns finite temperature and noise. At nonzero temperature, switching currents acquire a finite distribution rather than sharp thresholds, leading to a smearing of the fingerprints on a scale set by the noise energy relative to the Josephson energy $E_\mathrm{J}=\hbar I_\mathrm{c1}^{(0)}/(2e)$. For typical Al or Nb junctions considered in Sec.~\ref{sec:experiment}, $E_\mathrm{J}/k_\mathrm{B}\sim 1$--$10$~K, so operation in dilution refrigerators should keep thermal broadening below the characteristic feature contrast. Finally, we assume identical dynamical parameters $(\beta_\mathrm{c},R,C)$ for the two junctions. In practice, junction-to-junction variations will smooth nodal lines but are not expected to affect the qualitative discrimination between cases (a)–(d).
 
It is useful to place the proposed protocol in the context of alternative diagnostics of nontrivial CPR content. Shapiro-step spectroscopy under microwave irradiation is a direct probe of $4\pi$-periodicity associated with Majorana modes, but it requires dedicated microwave circuitry and impedance matching and is sensitive to step smearing. Static $I_\mathrm{c}(\phi_\mathrm{dc})$ measurements provide complementary information on the CPR shape but suffer from a basic inverse-problem ambiguity, as different CPR combinations can produce similar $I_\mathrm{c}(\phi_\mathrm{dc})$ curves~\cite{ModernAspects2017,askerzade_2015,Yerin_review}. The proposed ac-flux protocol occupies a useful intermediate position: it requires only standard dc SQUID instrumentation supplemented by an ac flux modulation line, yet accesses a two-dimensional parameter space in which different microscopic CPR components produce distinct geometric signatures.
 
Finally, extending the analysis to higher CPR harmonics such as $\sin(\varphi/3)$ or $\sin 3\varphi$, which can arise in multicomponent scenarios or in charge-$6e$ vestigial orders~\cite{Ge}, is straightforward and would lead to additional Bessel dressings $J_0(\xi_0/3)$ and $J_0(3\xi_0)$ with distinct nodal structures. The present framework is essentially classical, treating the SQUID phases as $c$-numbers; quantum extensions including phase fluctuations, charge dynamics, and environmental decoherence in the small-junction limit would broaden its applicability to qubit-scale SQUIDs and circuit-QED platforms.

\subsection{Possible experimental realization}
\label{sec:experiment}
The CPR identification protocol is as follows. One records $\eta(\phi_\mathrm{ac},\omega)$ on a representative grid in the parameter space and reads off the qualitative features (weak arcs, wide arcs, dense arcs, or
intermodulation patterns) to infer the underlying CPR content of the junctions. 
The protocol can be implemented using cryogenic setups for dc SQUIDs with conventional readout electronics. It relies on three experimental capabilities: a controllable dc magnetic flux for setting the operating point $\phi_\mathrm{dc}$, an ac flux modulation with tunable amplitude $\phi_\mathrm{ac}$ and frequency $\omega$, and four-probe current–voltage measurements enabling extraction of the switching critical currents $I_\mathrm{c}^{\pm}$. The dc and ac flux components can be delivered via a single on-chip flux line or an external coil.

The frequency range of interest, $\omega\sim 0.5$-$5$ in units of the plasma frequency, corresponds to frequencies in the GHz range for Al or Nb-based junctions, well within the bandwidth of standard cryogenic electronics \cite{Sterck1995,Dob20pra}. The ac flux amplitude range $\phi_\mathrm{ac}\sim 0.1$-$3$ in units of $\Phi_0/(2\pi)$ is similarly modest and routinely accessible. With the working point $(\phi_\mathrm{dc}=0.25,\,\alpha=0.7)$ used in this work, the predicted diode efficiencies $|\eta|$ exceed $0.1$ over significant portions of the $(\phi_\mathrm{ac},\omega)$ plane for both the subharmonic and second harmonic CPR cases, well above the typical sensitivity threshold of SQUID measurements.

Several material classes \cite{Dob26sst} constitute interesting platforms for these measurements. The first class includes asymmetric SQUIDs with high-transparency junctions or junctions based on topological materials, such as HgTe heterostructures, InAs quantum wells, and topological insulators, for which theory predicts subharmonic $\sin(\varphi/2)$ contributions~\cite{Cayao_Majorana,BoLu2022,BoLu2023,mondal2025}. Such a SQUID would exhibit the case-(b) fingerprint of widely spaced arcs determined by $J_0(\xi_0/2)$.

A second class consists of multiband and multicomponent superconductors, where multiple Fermi-surface sheets and interband phase structure naturally generate nonsinusoidal CPRs, including prominent $\sin 2\varphi$ contributions and, in frustrated regimes, $\pi$-junction behavior~\cite{Yerin_diode,Yerin_SQUID,Yerin2014,Yerin_review,Kiyko,askerzade_2015}. Iron-based superconductors, including FeSe, LiFeAs, and the 122-family pnictides, are representative multiband $s_\pm$ systems in which Josephson interferometry and proximity-effect measurements have already revealed substantial second-harmonic CPR contributions~\cite{Yerin_diode,Yerin_review}. Similar considerations apply to MgB$_2$~\cite{Yin_2024}, kagome superconductors~\cite{Chen2026}, and potentially nickelate systems~\cite{Koren2025}. In all these cases, the expected signature corresponds to the case-(c), and more generally the mixed case-(d), characterized by dense Bessel-arc patterns governed by $J_0(2\xi_0)$ and its intermodulation with $J_0(\xi_0/2)$.

Finally, there has been growing theoretical and experimental interest in vestigial-order phases supporting charge-$4e$ superconductivity, where a four-fermion bound state condenses while pair condensation is suppressed by fluctuations~\cite{Hecker2018,Ge}. The Josephson coupling between a conventional $2e$ superconductor and a charge-$4e$ condensate is $\pi$-periodic in the superconducting phase and corresponds, within our effective description, to a $\sin 2\varphi$ harmonic in Eq.~(\ref{eq:CPR2}). In an asymmetric SQUID, a charge-$4e$ condensate in one arm would therefore produce a clean case-(c) fingerprint, characterized by dense $J_0(2\xi_0)$-controlled arc structures in the $(\phi_\mathrm{ac},\omega)$ plane without subharmonic admixture. While charge-$4e$ superconductivity remains an active theoretical and experimental pursuit rather than an established material class, the proposed ac-flux protocol may nonetheless provide a transport-based diagnostic complementary to existing thermodynamic and spectroscopic probes.

\section{Conclusions}
\label{sec:conclusions}

We have introduced ac-flux-driven SQUID diode spectroscopy as a probe of nontrivial current–phase relations. The central result is that the two-dimensional phase diagram $\eta(\phi_\mathrm{ac},\omega)$ encodes qualitatively distinct fingerprints of the microscopic CPR content of the junctions.

Using a Kapitza-type analysis of the fast-driven dynamics [Eq.~(\ref{eq:eta_jacobi_anger_main})], we have shown that each CPR harmonic acquires a characteristic Bessel dressing. The conventional $\sin\varphi$ term is weighted by $J_0(\xi_0)$, the subharmonic $\sin(\varphi/2)$ by $J_0(\xi_0/2)$, and the second harmonic $\sin 2\varphi$ by $J_0(2\xi_0)$. As a result, the corresponding arcs exhibit distinct spacing patterns in the $(\phi_\mathrm{ac},\omega)$ plane: sparse for the subharmonic case, dense for the second harmonic case, and intermodulated in the mixed regime. By contrast, a conventional asymmetric SQUID with purely sinusoidal CPRs exhibits a much weaker response, governed by the static finite-inductance contribution $\eta_\mathrm{L}$ and an ac correction $\Delta\eta_\mathrm{ac}$. We derived this contribution using a Kapitza expansion and extended it to a nonperturbative form [Eq.~(\ref{eq:eta_a_nonpert})]. Even in this reference case, faint arcs governed by $J_0(\xi_0)$ appear, but they remain parametrically suppressed relative to the exotic CPR signatures.

These discriminating features are robust across dynamical regimes. We verified their persistence from the overdamped regime ($\beta_\mathrm{c}=0.1$) to the underdamped regime ($\beta_\mathrm{c}=3$), where $\beta_\mathrm{c}$ primarily affects arc curvature via the resonance position. The CPR scenarios are therefore distinguishable across several complementary observables. They yield distinct static fingerprints at $\phi_\mathrm{ac}=0$, different full phase diagrams in $\eta(\phi_\mathrm{ac},\omega)$, and clearly separated high-frequency limits in $\eta(\omega)$. They also produce qualitatively different drive-amplitude responses at fixed frequency. In particular, a single high-frequency measurement ($\omega \gtrsim 5,\tau_\mathrm{p}^{-1}$) or a two-frequency amplitude scan is sufficient to discriminate between all four cases. More generally, the full two-dimensional $\eta(\phi_\mathrm{ac},\omega)$ map provides a level of CPR resolution not accessible in single-frequency ac transport. This extends the approach of Ref.~\cite{Cuozzo2024} by incorporating frequency-resolved structure. As a result, ac-flux SQUID diode spectroscopy offers a practical transport-based diagnostic for nontrivial CPRs in asymmetric dc SQUIDs.

\section*{Supplemental Material}
Detailed analytical
derivations are provided in Supplemental Material \cite{SupplementalMaterial}.

\section*{Acknowledgements}
Y.Y. acknowledges CryoQuant/TU Braunschweig for hospitality and the Deutsche Forschungsgemeinschaft (DFG, German Research Foundation) under Germany's Excellence Strategy – EXC-2123 QuantumFrontiers – 390837967  for financial support. A.F. acknowledges the Department of Physics and Astronomy of the NASU for support through fundamental scientific program 0122U001501.
O.D. acknowledges the DFG for support through Grant No 577881064 (Super3DMag). The numerical calculations reported in this paper were partially performed at TUBITAK ULAKBIM, High Performance and Grid Computing Center (TRUBA resources). This research is based upon work from COST Action CA21144 (SuperQuMap) supported by the European Cooperation in Science and Technology.

\bibliography{REF}

\clearpage

\setcounter{equation}{0}
\setcounter{figure}{0}
\setcounter{section}{0}
\setcounter{table}{0}
\renewcommand{\theequation}{S\arabic{equation}}
\renewcommand{\thefigure}{S\arabic{figure}}
\renewcommand{\thesection}{S\Roman{section}}
\renewcommand{\thetable}{S\arabic{table}}

\begin{center}
{\large\bfseries Supplemental Material:\\[2pt]
AC-flux-driven SQUID diode spectroscopy as a probe of current-phase relations}
\end{center}
\vspace{1em}

\section{Analytical derivation of the diode efficiency}
\label{app:analytics}

We consider pure sinusoidal junctions with amplitude
asymmetry, $i_{c1}=\sin\varphi_1$ and
$i_{c2}=\alpha\sin\varphi_2$ with $\alpha\neq 1$. We
introduce the sum and difference variables:
\begin{equation}
\varphi_+=\frac{\varphi_1+\varphi_2}{2},
\qquad
\varphi_-=\frac{\varphi_1-\varphi_2}{2},
\label{eq:phi_pm}
\end{equation}
and the compact notation $a=(1+\alpha)/2$, $b=(1-\alpha)/2$,
with auxiliary functions:
\begin{align}
f(X,Y) &= a\sin X\cos Y + b\cos X\sin Y,
\label{eq:f_def}\\
g(X,Y) &= b\sin X\cos Y + a\cos X\sin Y.
\label{eq:g_def}
\end{align}

Adding and subtracting the equations of motion yields:
\begin{align}
\beta_c\ddot{\varphi}_+ + \dot{\varphi}_+
+ f(\varphi_+,\varphi_-)
= \frac{i_b}{2},
\label{eq:plus_compact}\\
\beta_c\ddot{\varphi}_- + \dot{\varphi}_-
+ g(\varphi_+,\varphi_-)
+ \frac{2(\varphi_- -\pi\phi_e)}{\pi\beta_L}= 0.
\label{eq:minus_compact}
\end{align}

In the limit $\beta_L\ll 1$, the difference phase is
approximately pinned: $\varphi_-\simeq\Psi\equiv\pi\phi_{dc}$.
The effective force on $\varphi_+$ reduces to:
\begin{equation}
f_0(X) = f(X,\Psi) = R_0\sin(X+\theta),
\label{eq:f0}
\end{equation}
with $R_0$ given by Eq.~(16) and:
\begin{equation}
\tan\theta =
\frac{(1-\alpha)\sin\pi\phi_{dc}}
{(1+\alpha)\cos\pi\phi_{dc}}.
\label{eq:theta_app}
\end{equation}
The leading-order effective equation is a shifted single-
harmonic washboard with $I_c^+=|I_c^-|$ and $\eta=0$.
The diode effect therefore requires corrections that generate
higher harmonics in the effective force.

Writing $\varphi_-(t)=\Psi+\xi(t)$ and retaining only fast-
varying terms in Eq.~(\ref{eq:minus_compact}):
\begin{equation}
\ddot{\xi}+\gamma\dot{\xi}+\omega_L^2\xi
= \frac{2\phi_{ac}}{\beta_c\beta_L}\cos\omega t,
\label{eq:fast_osc}
\end{equation}
with $\omega_L$ and $\gamma$ from Eq.~(17) in the main text.
The steady-state solution $\xi(t)=\xi_0\cos(\omega t-\delta)$
has amplitude and phase given by Eq.~(18) in the main text.
The perturbative treatment below assumes $\xi_0\ll 1$.

Averaging $f(\varphi_+,\Psi+\xi)$ over fast oscillations
gives:
\begin{equation}
\langle f(\varphi_+,\Psi+\xi)\rangle
= \left(1-\frac{\xi_0^2}{4}\right)f_0(\varphi_+)
+ O(\xi_0^4).
\label{eq:kapitza_renorm}
\end{equation}
This direct Kapitza correction merely re-normalizes the
amplitude $R_0\to R_1=(1-\xi_0^2/4)R_0$, producing no diode
effect.

For finite $\beta_L$, the slow component of $\varphi_-$ is
not exactly pinned. Solving Eq.~(\ref{eq:minus_compact})
quasi-statically to leading order in $\beta_L$ and using the
identity $g(X,\Psi)=-f_{XY}(X,\Psi)$:
\begin{equation}
\varphi_- = \Psi - \frac{\pi\beta_L}{2}g(X,\Psi)
+ O(\beta_L^2),
\label{eq:phi_minus_static}
\end{equation}

\begin{equation}
f(X,\varphi_-) \simeq f_0(X)
+ \frac{\pi\beta_L}{2}f_Y(X,\Psi)f_{XY}(X,\Psi),
\label{eq:static_correction}
\end{equation}
where the second term is a second harmonic in $X$:
\begin{align}
f_Y f_{XY} = & \frac{a^2\sin^2\!\Psi-b^2\cos^2\!\Psi}{2}
\sin 2X 
\nonumber\\
&
- ab\sin\Psi\cos\Psi\,\cos 2X.
\label{eq:fYfXY}
\end{align}
This finite-inductance second harmonic is the origin of
$\eta_L$ in Eq.~(14) in the main text.

The fast oscillation $\xi(t)$ drives a fast correction
$\zeta(t)$ of the common phase via $\varphi_+(t)=X(t)+\zeta(t)$. Linearizing Eq.~(\ref{eq:plus_compact}):
\begin{equation}
\beta_c\ddot{\zeta} + \dot{\zeta} + f_X(X,\Psi)\zeta
= -f_Y(X,\Psi)\xi(t),
\label{eq:zeta_eq}
\end{equation}
with common-mode susceptibility:
\begin{equation}
\chi_+(X,\omega)
= \frac{1}{f_X(X,\Psi)-\beta_c\omega^2+i\omega}.
\label{eq:chi}
\end{equation}

The averaged back-action contribution to the effective force
is:
\begin{equation}
\langle f_{XY}\zeta\xi\rangle
= -\frac{\xi_0^2}{2}\,\mathrm{Re}\,\chi_+(X,\omega)
\,f_Y f_{XY}.
\label{eq:backaction}
\end{equation}

At the switching points $X_\pm=\pm\pi/2-\theta$ where
$f_X(X_\pm,\Psi)=0$:
\begin{equation}
\mathrm{Re}\,\chi_+(X_\pm,\omega)
= -\frac{\beta_c}{1+\beta_c^2\omega^2},
\label{eq:Re_chi}
\end{equation}
giving the AC-induced second-harmonic coefficient:
\begin{equation}
K_{ac} = \frac{\beta_c\xi_0^2}{2(1+\beta_c^2\omega^2)}.
\label{eq:Kac}
\end{equation}

The total effective force near the switching points is:
\begin{equation}
f_\mathrm{eff}(X) = f_0(X)
+ K_\mathrm{tot}\,f_Y f_{XY},
\label{eq:f_eff}
\end{equation}
where $K_\mathrm{tot}=K_L+K_{ac}$ with
$K_L=\pi\beta_L/2$. Evaluating $f_Y f_{XY}$ at
$X=X_\pm$ using $a^2-b^2=\alpha$ and
$ab=(1-\alpha^2)/4$:
\begin{equation}
\left[f_Y f_{XY}\right]_{X=X_\pm}
= \frac{\alpha(1-\alpha^2)\sin 2\Psi}{8R_0^2}.
\label{eq:Hcritical}
\end{equation}

The switching critical currents are:
\begin{align}
I_c^+ &= 2\!\left[R_0 + K_\mathrm{tot}
\frac{\alpha(1-\alpha^2)\sin 2\Psi}{8R_0^2}\right],
\label{eq:Ic_plus}\\
|I_c^-| &= 2\!\left[R_0 - K_\mathrm{tot}
\frac{\alpha(1-\alpha^2)\sin 2\Psi}{8R_0^2}\right],
\label{eq:Ic_minus}
\end{align}
giving:
\begin{equation}
\eta \simeq K_\mathrm{tot}
\frac{\alpha(1-\alpha^2)\sin(2\pi\phi_{dc})}{8R_0^3}.
\label{eq:eta_Ktot}
\end{equation}

Substituting $K_L$ and $K_{ac}$ with $\xi_0$ from
Eq.~(18) (main text) recovers Eqs.~(13)-(15) of the main text.

The result~(\ref{eq:eta_Ktot}) satisfies all required
symmetry constraints. For symmetric junctions ($\alpha=1$),
$\alpha(1-\alpha^2)=0$, so $\eta=0$ for all parameters. At
the time-reversal symmetric flux values ($\phi_{dc}=0$ or
$\phi_{dc}=\frac{1}{2}$), $\sin(2\pi\phi_{dc})=0$, so
$\eta=0$. In the absence of AC drive ($\phi_{ac}=0$),
$\xi_0=0$ and hence $K_{ac}=0$ and $\Delta\eta_{ac}=0$,
while the static contribution $\eta_L$ from
$K_L=\pi\beta_L/2$ remains finite.

Finally, for zero inductance ($\beta_L\to 0$) we get both $K_L\to 0$ and $\omega_L\to\infty$ so $\xi_0\to 0$ and $K_{ac}\to 0$.
Both contributions vanish and $\eta=0$, confirming that the
purely pinned dc SQUID has no diode effect from a shifted first
harmonic alone.

\subsection{Non-perturbative dressing of the induced diode efficiency}
\label{app:nonpert}

Eqs.~(13)--(15) and \eqref{eq:eta_Ktot} were obtained by expanding the fast average to leading order in $\xi_0$
[cf.\ Eq.~(\ref{eq:kapitza_renorm})]. We now retain the fast
oscillation $\xi(t)=\xi_0\cos(\omega t-\delta)$ exactly inside
the trigonometric functions, which extends the result to
arbitrary $\xi_0$ within the pinned-flux approximation and
explains the faint arc structure observed in the conventional
case at high numerical resolution [Fig.~6(a)].

The diode effect originates from the second-harmonic
content of the effective force generated by the
finite-$\beta_L$ feedback and the AC back-action,
\begin{equation}
H(X)\equiv f_Y(X,\varphi_-)\,f_{XY}(X,\varphi_-),
\label{eq:Hdef_np}
\end{equation}
which in the leading-order treatment was evaluated at the
pinned value $\varphi_-=\Psi$, Eq.~(\ref{eq:fYfXY}). When the
difference phase carries the full fast modulation
$\varphi_-=\Psi+\xi(t)$, the product $H$ must be averaged over
the fast cycle. Writing $f_Y$ and $f_{XY}$ explicitly,
\begin{align}
f_Y(X,Y) &= -a\sin X\sin Y + b\cos X\cos Y,
\label{eq:fY_np}\\
f_{XY}(X,Y) &= -a\cos X\sin Y - b\sin X\cos Y,
\label{eq:fXY_np}
\end{align}
their product is quadratic in $\{\cos Y,\sin Y\}$ and
therefore contains the second harmonic $\cos 2Y$, $\sin 2Y$,
together with a $Y$-independent part. Using
$\langle\cos 2Y\rangle=J_0(2\xi_0)\cos 2\Psi$ and
$\langle\sin 2Y\rangle=J_0(2\xi_0)\sin 2\Psi$ for
$Y=\Psi+\xi_0\cos(\omega t-\delta)$, the fast average of $H$
splits into an undressed constant piece and a $J_0(2\xi_0)$-dressed
oscillating piece. Evaluating the result at the switching
extrema $X_\pm=\pm\pi/2-\theta$ and summing the two
contributions [as required by Eqs.~(\ref{eq:Ic_plus})--(\ref{eq:Ic_minus})],
a short calculation gives
\begin{equation}
\bigl\langle H\bigr\rangle_{X_+}
+\bigl\langle H\bigr\rangle_{X_-}
=
\bigl[1+J_0(2\xi_0)\bigr]\,
\frac{\alpha(1-\alpha^2)\sin 2\Psi}{8R_0^2},
\label{eq:H_extrema_np}
\end{equation}
which reduces to the leading-order value
$\alpha(1-\alpha^2)\sin 2\Psi/(4R_0^2)$
[twice Eq.~(\ref{eq:Hcritical})] when $\xi_0\to 0$ and
$J_0(2\xi_0)\to 1$.

The dressed first harmonic, on the other hand, sets the
denominator of the diode efficiency through
$I_1=J_0(\xi_0)R_0$ [Eq.~(\ref{eq:kapitza_renorm}) with the
exact factor $J_0(\xi_0)$ in place of $1-\xi_0^2/4$].
Combining the second-harmonic numerator
Eq.~(\ref{eq:H_extrema_np}) with this denominator,
\begin{equation}
I_c^\pm = 2\Bigl[J_0(\xi_0)R_0 \pm K_{\mathrm{tot}}\,
\tfrac{1}{2}\bigl(\langle H\rangle_{X_+}+\langle H\rangle_{X_-}\bigr)\Bigr],
\label{eq:Ic_np}
\end{equation}
we obtain the non-perturbative induced diode efficiency
\begin{equation}
\eta^{(a)}
\simeq
\frac{K_{\mathrm{tot}}\bigl[1+J_0(2\xi_0)\bigr]\,
\alpha(1-\alpha^2)\sin(2\pi\phi_{dc})}
{16\,J_0(\xi_0)\,R_0^3},
\label{eq:eta_a_nonpert_app}
\end{equation}
with $K_{\mathrm{tot}}=\pi\beta_L/2+\beta_c\xi_0^2/[2(1+\beta_c^2\omega^2)]$
as in Eq.~(\ref{eq:Ktot_main}).

Two limits confirm its consistency. For $\xi_0\ll 1$, the
ratio $[1+J_0(2\xi_0)]/[2J_0(\xi_0)]\to 1$ and
Eq.~(\ref{eq:eta_a_nonpert_app}) reduces exactly to
$\eta_L+\Delta\eta_{ac}$, Eqs.~(13)-(15) of the main text.
For larger drive, the denominator $J_0(\xi_0)R_0$ vanishes
whenever $\xi_0$ reaches a zero $j_{0,n}$ of $J_0$; the
expansion Eq.~(\ref{eq:Ic_np}) then ceases to be controlled
(the neglected subleading harmonics become comparable to the
collapsing first harmonic), and the diode efficiency
saturates rather than diverges. Physically, the loci
$\xi_0(\phi_{ac},\omega)=j_{0,n}$ mark the lines where the
dressed first harmonic is weakest and the residual induced
second harmonic dominates the effective CPR, producing the
faint arc-like enhancements of the conventional case. Their
spacing is set by the zeros of $J_0(\xi_0)$, intermediate
between the $J_0(\xi_0/2)$ spacing of the subharmonic case
and the $J_0(2\xi_0)$ spacing of the second-harmonic case. The arcs are, however, strongly suppressed in amplitude by the prefactor $K_{\mathrm{tot}}$, which is of order $\beta_L$ at low drive, so that the conventional case remains the weak reference
response throughout.

\section{Derivation of the effective averaged CPR and diode efficiency}
\label{app:jacobi_anger}

In this section we derive the effective CPR and the corresponding leading-order diode efficiency in the presence of a general CPR of the form
\begin{align}
i_{c1}(\varphi_1) &= \sin\varphi_1,
\label{eq:app_CPR1}\\
i_{c2}(\varphi_2) &= \alpha\sin\varphi_2
+ i_{c2}^{(1)}\sin\!\left(\frac{\varphi_2}{2}\right)
+ i_{c2}^{(2)}\sin 2\varphi_2.
\label{eq:app_CPR2}
\end{align}
Our goal is to obtain an analytical expression for the diode efficiency \(\eta\) in the fast-averaged regime using the Jacobi--Anger expansion. The derivation is controlled in the limit of strong inductive screening, \(\beta_L\ll 1\), where the difference phase \(\varphi_-\) remains close to the externally imposed flux, and the fast AC modulation can be averaged over one drive period.

We begin from the equations of motion
\begin{align}
\beta_c \ddot{\varphi}_1 + \dot{\varphi}_1
+ i_{c1}(\varphi_1) &= \frac{i_b}{2} - j,
\label{eq:app_EOM1}\\
\beta_c \ddot{\varphi}_2 + \dot{\varphi}_2
+ i_{c2}(\varphi_2) &= \frac{i_b}{2} + j,
\label{eq:app_EOM2}
\end{align}
with
\begin{equation}
j=\frac{\varphi_1-\varphi_2-2\pi\phi_e}{\pi\beta_L},
\quad
\phi_e(t)=\phi_{dc}+\phi_{ac}\cos\omega t .
\label{eq:app_flux}
\end{equation}
We introduce the standard sum and difference variables
\begin{equation}
\varphi_+=\frac{\varphi_1+\varphi_2}{2},
\qquad
\varphi_-=\frac{\varphi_1-\varphi_2}{2},
\label{eq:app_phi_pm}
\end{equation}
so that
\begin{equation}
\varphi_1=\varphi_+ + \varphi_-,
\qquad
\varphi_2=\varphi_+ - \varphi_- .
\label{eq:app_phi12}
\end{equation}

Adding Eqs.~\eqref{eq:app_EOM1} and \eqref{eq:app_EOM2} and dividing by
\(2\), we obtain the equation for the common phase:
\begin{equation}
\beta_c\ddot{\varphi}_+ + \dot{\varphi}_+
+\frac12\Bigl[i_{c1}(\varphi_+ + \varphi_-)
+i_{c2}(\varphi_+ - \varphi_-)\Bigr]
=
\frac{i_b}{2}.
\label{eq:app_plus_eq_general}
\end{equation}
Substituting Eqs.~\eqref{eq:app_CPR1} and \eqref{eq:app_CPR2}, this becomes
\begin{align}
&\beta_c\ddot{\varphi}_+ + \dot{\varphi}_+
+\frac12\sin(\varphi_+ + \varphi_-)
+\frac{\alpha}{2}\sin(\varphi_+ - \varphi_-)
\nonumber\\
&
+\frac{i_{c2}^{(1)}}{2}\sin\!\left(\frac{\varphi_+ - \varphi_-}{2}\right)
\nonumber\\
&
+\frac{i_{c2}^{(2)}}{2}\sin\!\bigl[2(\varphi_+ - \varphi_-)\bigr]
=
\frac{i_b}{2}.
\label{eq:app_plus_eq_expanded}
\end{align}

Similarly, subtracting Eq.~\eqref{eq:app_EOM2} from
Eq.~\eqref{eq:app_EOM1} and dividing by \(2\), one finds
\begin{align}
&\beta_c\ddot{\varphi}_- + \dot{\varphi}_-
+\frac12\sin(\varphi_+ + \varphi_-)
-\frac{\alpha}{2}\sin(\varphi_+ - \varphi_-)
\nonumber\\
&
-\frac{i_{c2}^{(1)}}{2}\sin\!\left(\frac{\varphi_+ - \varphi_-}{2}\right)
-\frac{i_{c2}^{(2)}}{2}\sin\!\bigl[2(\varphi_+ - \varphi_-)\bigr]
\nonumber\\
&
+\frac{2(\varphi_- - \pi\phi_e)}{\pi\beta_L}
=0.
\label{eq:app_minus_eq_expanded}
\end{align}

We now specialize to the regime
\begin{equation}
\beta_L\ll 1.
\label{eq:app_small_betaL}
\end{equation}
In this limit, the inductive term strongly locks the difference phase
to the applied flux. We therefore write
\begin{equation}
\varphi_-(t)=\Psi+\xi(t),
\qquad
\Psi=\pi\phi_{dc},
\label{eq:app_phi_minus_ansatz}
\end{equation}
where \(\Psi\) is the slow DC-flux part and \(\xi(t)\) is the fast
oscillation induced by the AC drive.

To leading order, the fast dynamics of \(\xi(t)\) is dominated by the inductive restoring force and the AC flux drive. Neglecting the
Josephson terms in Eq.~\eqref{eq:app_minus_eq_expanded} at this stage, the fast component obeys
\begin{equation}
\beta_c\ddot{\xi}+\dot{\xi}+\frac{2}{\pi\beta_L}\xi
=
\frac{2\phi_{ac}}{\beta_L}\cos\omega t .
\label{eq:app_xi_eq}
\end{equation}
Equivalently,
\begin{equation}
\ddot{\xi}+\gamma\dot{\xi}+\omega_L^2\xi
=
\frac{2\phi_{ac}}{\beta_c\beta_L}\cos\omega t,
\label{eq:app_xi_eq_std}
\end{equation}
with
\begin{equation}
\omega_L^2=\frac{2}{\pi\beta_c\beta_L},
\quad
\gamma=\frac{1}{\beta_c}.
\label{eq:app_omegaL_gamma}
\end{equation}
The steady-state solution is
\begin{equation}
\xi(t)=\xi_0\cos(\omega t-\delta),
\label{eq:app_xi_sol}
\end{equation}
where
\begin{equation}
\xi_0=
\frac{2\phi_{ac}/(\beta_c\beta_L)}
{\sqrt{(\omega_L^2-\omega^2)^2+\gamma^2\omega^2}},
\quad
\tan\delta=
\frac{\gamma\omega}{\omega_L^2-\omega^2}.
\label{eq:app_xi0_delta}
\end{equation}

We work in the fast-averaging regime in which
\(\xi(t)\) is retained exactly inside trigonometric functions, while
its slow back-action on the \(\varphi_-\) dynamics is neglected at
leading order. This is the natural regime for a Jacobi--Anger treatment.

The key identity is the Jacobi--Anger expansion
\begin{equation}
e^{iz\cos\theta}
=
\sum_{m=-\infty}^{\infty} i^m J_m(z)e^{im\theta},
\label{eq:app_JA}
\end{equation}
where \(J_m(z)\) is the Bessel function of the first kind.
Taking the imaginary part of
\begin{equation}
e^{i[A+z\cos\theta]}=e^{iA}e^{iz\cos\theta},
\end{equation}
we obtain
\begin{equation}
\sin\!\bigl(A+z\cos\theta\bigr)
=
\sum_{m=-\infty}^{\infty}
J_m(z)\,
\sin\!\left(A+m\theta+\frac{m\pi}{2}\right).
\label{eq:app_sin_JA}
\end{equation}
Averaging over one period of the fast oscillation,
\begin{equation}
\langle \cdots \rangle
=
\frac{\omega}{2\pi}\int_0^{2\pi/\omega}(\cdots)\,dt,
\label{eq:app_time_average}
\end{equation}
eliminates all harmonics with \(m\neq 0\), so that only the
\(m=0\) term survives:
\begin{equation}
\left\langle \sin\!\bigl(A+z\cos(\omega t-\delta)\bigr)\right\rangle
=
J_0(z)\sin A.
\label{eq:app_basic_average}
\end{equation}
Since \(J_0(z)\) is an even function, the same result holds for
\(\sin(A-z\cos(\omega t-\delta))\).

Eq. ~\eqref{eq:app_basic_average} is the basic reason why each CPR
harmonic acquires its own Bessel dressing factor. The argument of the
Bessel function is proportional to the phase-modulation amplitude
experienced by that harmonic.

Substituting
\begin{equation}
\varphi_-(t)=\Psi+\xi_0\cos(\omega t-\delta)
\label{eq:app_phi_minus_modulated}
\end{equation}
into Eq.~\eqref{eq:app_plus_eq_expanded}, the four supercurrent terms become
\begin{align}
\frac12\sin(\varphi_+ + \varphi_-)
&=
\frac12\sin\!\bigl(\varphi_+ + \Psi + \xi(t)\bigr),
\label{eq:app_term1}
\\
\frac{\alpha}{2}\sin(\varphi_+ - \varphi_-)
&=
\frac{\alpha}{2}\sin\!\bigl(\varphi_+ - \Psi - \xi(t)\bigr),
\label{eq:app_term2}
\\
\frac{i_{c2}^{(1)}}{2}\sin\!\left(\frac{\varphi_+ - \varphi_-}{2}\right)
&=
\frac{i_{c2}^{(1)}}{2}
\sin\!\left(
\frac{\varphi_+ - \Psi}{2}
-\frac{\xi(t)}{2}
\right),
\label{eq:app_term3}
\\
\frac{i_{c2}^{(2)}}{2}\sin\!\bigl[2(\varphi_+ - \varphi_-)\bigr]
&=
\frac{i_{c2}^{(2)}}{2}
\sin\!\bigl(
2\varphi_+ -2\Psi -2\xi(t)
\bigr).
\label{eq:app_term4}
\end{align}

We now average each term separately.

\paragraph{Ordinary first harmonic from junction 1.}
Using Eq.~\eqref{eq:app_basic_average} with
\(A=\varphi_+ + \Psi\) and \(z=\xi_0\), we obtain
\begin{equation}
\left\langle
\frac12\sin(\varphi_+ + \Psi + \xi)
\right\rangle
=
\frac12 J_0(\xi_0)\sin(\varphi_+ + \Psi).
\label{eq:app_avg_term1}
\end{equation}

\paragraph{Ordinary first harmonic from junction 2.}
Again using Eq.~\eqref{eq:app_basic_average},
\begin{equation}
\left\langle
\frac{\alpha}{2}\sin(\varphi_+ - \Psi - \xi)
\right\rangle
=
\frac{\alpha}{2}J_0(\xi_0)\sin(\varphi_+ - \Psi).
\label{eq:app_avg_term2}
\end{equation}

\paragraph{Fractional harmonic \(\sin(\varphi_2/2)\).}
Now the phase modulation enters with half the amplitude,
\(z=\xi_0/2\), so
\begin{equation}
\left\langle
\frac{i_{c2}^{(1)}}{2}
\sin\!\left(
\frac{\varphi_+ - \Psi}{2}-\frac{\xi}{2}
\right)
\right\rangle
=
\frac{i_{c2}^{(1)}}{2}
J_0\!\left(\frac{\xi_0}{2}\right)
\sin\!\left(\frac{\varphi_+ - \Psi}{2}\right).
\label{eq:app_avg_term3}
\end{equation}

\paragraph{Second harmonic \(\sin 2\varphi_2\).}
Here the phase modulation enters with twice the amplitude,
\(z=2\xi_0\), so
\begin{equation}
\left\langle
\frac{i_{c2}^{(2)}}{2}
\sin\!\bigl(2\varphi_+ -2\Psi -2\xi\bigr)
\right\rangle
=
\frac{i_{c2}^{(2)}}{2}
J_0(2\xi_0)\sin\!\bigl(2\varphi_+ -2\Psi\bigr).
\label{eq:app_avg_term4}
\end{equation}

Combining Eqs.~\eqref{eq:app_avg_term1}--\eqref{eq:app_avg_term4}, the
fast-averaged common-phase equation takes the form
\begin{equation}
\beta_c\ddot{\varphi}_+ + \dot{\varphi}_+
+ I_{\mathrm{eff}}(\varphi_+)
=
\frac{i_b}{2},
\label{eq:app_reduced_eq}
\end{equation}
with the effective averaged CPR
\begin{align}
I_{\mathrm{eff}}(\varphi_+)
&=
\frac12 J_0(\xi_0)\sin(\varphi_+ + \Psi)
\nonumber\\
&\quad
+\frac{\alpha}{2}J_0(\xi_0)\sin(\varphi_+ - \Psi)
\nonumber\\
&\quad
+\frac{i_{c2}^{(1)}}{2}J_0\!\left(\frac{\xi_0}{2}\right)
\sin\!\left(\frac{\varphi_+ - \Psi}{2}\right)
\nonumber\\
&\quad
+\frac{i_{c2}^{(2)}}{2}J_0(2\xi_0)\sin\!\bigl(2\varphi_+ -2\Psi\bigr).
\label{eq:app_Ieff_raw}
\end{align}

The first two terms in Eq.~\eqref{eq:app_Ieff_raw} may be combined into a single shifted first harmonic. Using
\begin{align}
&\frac12\sin(\varphi_+ + \Psi)
+\frac{\alpha}{2}\sin(\varphi_+ - \Psi) =
\nonumber\\
&\quad
\frac{1+\alpha}{2}\sin\varphi_+\cos\Psi
+\frac{1-\alpha}{2}\cos\varphi_+\sin\Psi,
\label{eq:app_firstharm_expand}
\end{align}
we define
\begin{equation}
a=\frac{1+\alpha}{2},
\qquad
b=\frac{1-\alpha}{2},
\label{eq:app_ab}
\end{equation}
and
\begin{equation}
R_0=
\sqrt{
a^2\cos^2\Psi+b^2\sin^2\Psi
},
\label{eq:app_R0}
\end{equation}
together with the phase shift
\begin{equation}
\tan\theta=\frac{b\sin\Psi}{a\cos\Psi}
=
\frac{(1-\alpha)\sin\Psi}{(1+\alpha)\cos\Psi}.
\label{eq:app_theta}
\end{equation}
Then
\begin{equation}
\frac12\sin(\varphi_+ + \Psi)
+\frac{\alpha}{2}\sin(\varphi_+ - \Psi)
=
R_0\sin(\varphi_+ + \theta),
\label{eq:app_R0theta_identity}
\end{equation}
hence, the effective CPR becomes
\begin{equation}
\begin{aligned}
I_{\mathrm{eff}}(\varphi_+) =\;&
I_1\sin(\varphi_+ + \theta)
+ I_{1/2}\sin\!\left(\frac{\varphi_+}{2}+\chi_{1/2}\right)\\
&+ I_2\sin(2\varphi_+ + \chi_2),
\end{aligned}
\label{eq:app_Ieff_compact}
\end{equation}
with
\begin{align}
I_1 &= J_0(\xi_0)\,R_0,
\label{eq:app_I1}\\
I_{1/2} &= \frac{i_{c2}^{(1)}}{2}J_0\!\left(\frac{\xi_0}{2}\right),
\label{eq:app_Ihalf}\\
I_2 &= \frac{i_{c2}^{(2)}}{2}J_0(2\xi_0),
\label{eq:app_I2}
\end{align}
and phase offsets
\begin{align}
\chi_{1/2} &= -\frac{\Psi}{2},
\label{eq:app_chi_half}\\
\chi_2 &= -2\Psi.
\label{eq:app_chi2}
\end{align}

Eq.~\eqref{eq:app_Ieff_compact} is the central result of the
Jacobi--Anger reduction: each CPR harmonic is dressed by a different
Bessel factor,
\begin{align}
&\sin\varphi \ \rightarrow\ J_0(\xi_0),
\nonumber\\
&
\sin\!\left(\frac{\varphi}{2}\right)\ \rightarrow\ 
J_0\!\left(\frac{\xi_0}{2}\right),
\nonumber\\
&
\sin 2\varphi\ \rightarrow\ J_0(2\xi_0).
\label{eq:app_Bessel_dressing_summary}
\end{align}
This immediately explains why different microscopic CPR components
generate different arc spacings in the two-dimensional map
\(\eta(\phi_{ac},\omega)\): the zeros of the corresponding Bessel
functions occur at different values of the effective phase-modulation
amplitude.

We now assume that the dressed first harmonic dominates,
\begin{equation}
|I_{1/2}|,\ |I_2| \ll |I_1|,
\label{eq:app_perturbative_condition}
\end{equation}
and that we stay away from the lines where \(I_1=J_0(\xi_0)R_0\) becomes
anomalously small. In this regime, the extrema of the effective CPR are
close to those of the leading term \(I_1\sin(\varphi_+ + \theta)\):
\begin{equation}
\varphi_+^{(+)}=\frac{\pi}{2}-\theta,
\qquad
\varphi_+^{(-)}=-\frac{\pi}{2}-\theta.
\label{eq:app_extrema0}
\end{equation}

Let us write
\begin{equation}
I_{\mathrm{eff}}(\varphi_+)=I_1\sin(\varphi_+ + \theta)+g(\varphi_+),
\label{eq:app_decompose_g}
\end{equation}
where
\begin{equation}
g(\varphi_+)
=
I_{1/2}\sin\!\left(\frac{\varphi_+}{2}+\chi_{1/2}\right)
+
I_2\sin(2\varphi_+ + \chi_2).
\label{eq:app_g_def}
\end{equation}
To first order in \(g\), the shift of the extrema does not contribute
to the critical current, because the derivative of the dominant first
harmonic vanishes at \(\varphi_+^{(\pm)}\). Therefore
\begin{align}
I_c^+ &\simeq 2\Bigl[I_1+g(\varphi_+^{(+)})\Bigr],
\label{eq:app_Icp}\\
|I_c^-| &\simeq 2\Bigl[I_1-g(\varphi_+^{(-)})\Bigr].
\label{eq:app_Icm}
\end{align}
Consequently,
\begin{equation}
\eta
=
\frac{I_c^+-|I_c^-|}{I_c^+ + |I_c^-|}
\simeq
\frac{g(\varphi_+^{(+)})+g(\varphi_+^{(-)})}{2I_1}.
\label{eq:app_eta_general_g}
\end{equation}

We now evaluate the two contributions separately. For the fractional term,
\begin{equation}
g_{1/2}(\varphi_+)=
I_{1/2}\sin\!\left(\frac{\varphi_+}{2}+\chi_{1/2}\right),
\label{eq:app_g_half}
\end{equation}
we have
\begin{align}
g_{1/2}(\varphi_+^{(+)})
&=
I_{1/2}
\sin\!\left(
\frac{\pi}{4}-\frac{\theta}{2}+\chi_{1/2}
\right),
\label{eq:app_g_half_plus}
\\
g_{1/2}(\varphi_+^{(-)})
&=
I_{1/2}
\sin\!\left(
-\frac{\pi}{4}-\frac{\theta}{2}+\chi_{1/2}
\right).
\label{eq:app_g_half_minus}
\end{align}
Using \(\sin u+\sin v=2\sin[(u+v)/2]\cos[(u-v)/2]\), we obtain
\begin{align}
g_{1/2}(\varphi_+^{(+)})+g_{1/2}(\varphi_+^{(-)})
&=
-\sqrt{2}\,I_{1/2}
\sin\!\left(\frac{\Psi+\theta}{2}\right),
\label{eq:app_sum_half}
\end{align}
because \(\chi_{1/2}=-\Psi/2\).
Therefore the fractional contribution to the diode efficiency is
\begin{equation}
\eta_{1/2}
=
-\frac{I_{1/2}}{\sqrt{2}\,I_1}
\sin\!\left(\frac{\Psi+\theta}{2}\right).
\label{eq:app_eta_half_intermediate}
\end{equation}
Substituting Eqs.~\eqref{eq:app_I1} and \eqref{eq:app_Ihalf},
\begin{equation}
\eta_{1/2}
=
-\frac{
i_{c2}^{(1)}\,J_0(\xi_0/2)
}{
2\sqrt{2}\,J_0(\xi_0)\,R_0
}
\sin\!\left(\frac{\Psi+\theta}{2}\right).
\label{eq:app_eta_half_final}
\end{equation}

For the second harmonic,
\begin{equation}
g_2(\varphi_+)=I_2\sin(2\varphi_+ + \chi_2),
\label{eq:app_g2_def}
\end{equation}
we obtain
\begin{align}
g_2(\varphi_+^{(+)})
&=
I_2\sin\!\bigl(\pi-2\theta+\chi_2\bigr),
\label{eq:app_g2_plus}
\\
g_2(\varphi_+^{(-)})
&=
I_2\sin\!\bigl(-\pi-2\theta+\chi_2\bigr).
\label{eq:app_g2_minus}
\end{align}
Since \(\chi_2=-2\Psi\), one finds
\begin{equation}
g_2(\varphi_+^{(+)})+g_2(\varphi_+^{(-)})
=
2I_2\sin\!\bigl[2(\Psi+\theta)\bigr].
\label{eq:app_sum_g2}
\end{equation}
Hence the second-harmonic contribution to \(\eta\) is
\begin{equation}
\eta_2
=
\frac{I_2}{I_1}\sin\!\bigl[2(\Psi+\theta)\bigr].
\label{eq:app_eta2_intermediate}
\end{equation}
Substituting Eqs.~\eqref{eq:app_I1} and \eqref{eq:app_I2},
\begin{equation}
\eta_2
=
\frac{
i_{c2}^{(2)}\,J_0(2\xi_0)
}{
2\,J_0(\xi_0)\,R_0
}
\sin\!\bigl[2(\Psi+\theta)\bigr].
\label{eq:app_eta2_final}
\end{equation}

\subsection{Final perturbative result}

Adding Eqs.~\eqref{eq:app_eta_half_final} and
\eqref{eq:app_eta2_final}, we obtain the leading-order diode
efficiency in the fast-averaged pinned-flux regime:
\begin{align}
&\eta
\simeq
-\frac{
i_{c2}^{(1)}\,J_0(\xi_0/2)
}{
2\sqrt{2}\,J_0(\xi_0)\,R_0
}
\sin\!\left(\frac{\Psi+\theta}{2}\right)
+
\nonumber\\
&
\frac{
i_{c2}^{(2)}\,J_0(2\xi_0)
}{
2\,J_0(\xi_0)\,R_0
}
\sin\!\bigl[2(\Psi+\theta)\bigr].
\label{eq:app_eta_final}
\end{align}
Here
\begin{align}
\Psi &= \pi\phi_{dc},
\label{eq:app_Psi_again}\\
R_0&=
\sqrt{
\left(\frac{1+\alpha}{2}\right)^2\cos^2\Psi
+
\left(\frac{1-\alpha}{2}\right)^2\sin^2\Psi
},
\label{eq:app_R0_again}\\
\tan\theta&=
\frac{(1-\alpha)\sin\Psi}{(1+\alpha)\cos\Psi},
\label{eq:app_theta_again}\\
\xi_0&=
\frac{2\phi_{ac}/(\beta_c\beta_L)}
{\sqrt{(\omega_L^2-\omega^2)^2+\gamma^2\omega^2}},
\omega_L^2=\frac{2}{\pi\beta_c\beta_L},
\gamma=\frac{1}{\beta_c}.
\label{eq:app_xi0_again}
\end{align}

Equation~\eqref{eq:app_eta_final} shows explicitly that the fractional term \(\sin(\varphi/2)\) is dressed by \(J_0(\xi_0/2)\);
the ordinary first harmonic is dressed by \(J_0(\xi_0)\);
the second harmonic \(\sin 2\varphi\) is dressed by \(J_0(2\xi_0)\).
Therefore, the arc-like nodal patterns in the map
\(\eta(\phi_{ac},\omega)\) are expected to occur at different
locations depending on the microscopic CPR content.

\subsection{Limiting cases}

It is useful to derive several special cases of
Eq.~\eqref{eq:app_eta_final}.

\paragraph{Fractional term only.}
If \(i_{c2}^{(2)}=0\), then
\begin{equation}
\eta
\simeq
-\frac{
i_{c2}^{(1)}\,J_0(\xi_0/2)
}{
2\sqrt{2}\,J_0(\xi_0)\,R_0
}
\sin\!\left(\frac{\Psi+\theta}{2}\right).
\label{eq:app_eta_fractional_only}
\end{equation}
In this case the dominant zeros are controlled by
\begin{equation}
J_0(\xi_0/2)=0.
\label{eq:app_fractional_zero}
\end{equation}

\paragraph{Second harmonic only.}
If \(i_{c2}^{(1)}=0\), then
\begin{equation}
\eta
\simeq
\frac{
i_{c2}^{(2)}\,J_0(2\xi_0)
}{
2\,J_0(\xi_0)\,R_0
}
\sin\!\bigl[2(\Psi+\theta)\bigr].
\label{eq:app_eta_second_only}
\end{equation}
Now the dominant zeros are controlled by
\begin{equation}
J_0(2\xi_0)=0.
\label{eq:app_second_zero}
\end{equation}

\paragraph{Pure amplitude asymmetry.}
If \(i_{c2}^{(1)}=i_{c2}^{(2)}=0\), then
Eq.~\eqref{eq:app_eta_final} gives
\begin{equation}
\eta=0
\label{eq:app_eta_zero_pure_asym}
\end{equation}
in the strict pinned-flux fast-averaged approximation. This is
consistent with the fact that the reduced CPR then contains only a
shifted first harmonic. Therefore, a nonzero diode effect in the purely sinusoidal asymmetric case requires the next-order mechanisms discussed in the main text and derived in above, namely finite-\(\beta_L\) feedback and AC-induced back-action beyond the present leading Jacobi-Anger reduction.

The derivation presented here relies on the assumptions that
\(\beta_L\ll 1\), so that the difference phase remains close to the externally imposed flux; the fast mode \(\xi(t)\) is well described by Eq.~\eqref{eq:app_xi_sol};  the dressed first harmonic dominates, Eq.~\eqref{eq:app_perturbative_condition};
we stay away from lines where \(J_0(\xi_0)R_0\) becomes very
small, since the perturbative expansion for the critical currents then breaks down.

Near the nodal lines of \(J_0(\xi_0)\) or near flux values where
\(R_0\to 0\), the full effective CPR
\(I_{\mathrm{eff}}(\varphi_+)\) should be analyzed without expanding in
the subleading harmonics, and the critical currents must in general be determined numerically.

\subsection{Combined asymptotic formula: induced and direct diode contributions}
\label{app:combined_asymptotic}

In this subsection we combine the two complementary analytical
reductions developed in previous sections showed that,
for a purely sinusoidal asymmetric SQUID, a nonzero diode effect is generated only at next order through finite-$\beta_L$ feedback and AC-induced back-action. Section \ref{app:jacobi_anger} showed that, for a general CPR, explicit subharmonic and second-harmonic terms produce a diode effect already at the level of the leading fast-averaged effective CPR through their Jacobi-Anger Bessel dressings. The purpose of the present consideration is to derive a single asymptotic expression that combines these two mechanisms in the regime where they are simultaneously small.

The derivation below is controlled under assumptions of strong inductive screening:
\begin{equation}
\beta_L \ll 1,
\end{equation}
so that the difference phase remains close to the externally imposed
flux,
\begin{equation}
\phi_-(t)=\Psi+\xi(t)+\delta\phi_-(t),
\qquad
\Psi=\pi\phi_{dc},
\label{eq:app_comb_phi_minus}
\end{equation}
with \(\xi(t)\) the fast AC-driven oscillation and
\(\delta\phi_-(t)=O(\beta_L)\) the slower feedback correction.

Then we take into account fast-mode reduction, when the fast oscillation \(\xi(t)\) is governed, to leading order, by the
same driven linear oscillator as in Section \ref{app:jacobi_anger},
\begin{equation}
\beta_c\ddot{\xi}+\dot{\xi}+\frac{2}{\pi\beta_L}\xi
=
\frac{2\phi_{ac}}{\beta_L}\cos\omega t,
\label{eq:app_comb_xi_eq}
\end{equation}
with steady-state amplitude
\begin{equation}
\xi_0=
\frac{2\phi_{ac}/(\beta_c\beta_L)}
{\sqrt{(\omega_L^2-\omega^2)^2+\gamma^2\omega^2}},
\omega_L^2=\frac{2}{\pi\beta_c\beta_L},
\gamma=\frac{1}{\beta_c}.
\label{eq:app_comb_xi0}
\end{equation}

The effective first harmonic remains the largest term in the averaged CPR, namely
\begin{equation}
|I_{1/2}|,\ |I_2|,\ |I_2^{(\mathrm{ind})}| \ll |I_1|,
\label{eq:app_comb_dom}
\end{equation}
and we stay away from the nodal lines where
\begin{equation}
I_1=J_0(\xi_0)R_0 \approx 0.
\label{eq:app_comb_nodal}
\end{equation}
we retain the finite-$\beta_L$ and back-action correction generated by the purely sinusoidal sector, but neglect mixed terms such as
\begin{equation}
O\!\left(\beta_L i_{c2}^{(1)}\right),\quad
O\!\left(\beta_L i_{c2}^{(2)}\right),\quad
O\!\left(\xi_0^2 i_{c2}^{(1)}\right),\quad
O\!\left(\xi_0^2 i_{c2}^{(2)}\right),
\label{eq:app_comb_mixed}
\end{equation}
as well as products of the subharmonic and second-harmonic amplitudes.
Thus, the resulting expression is asymptotic rather than exact.

From Section \ref{app:jacobi_anger}, the fast-averaged CPR in the pinned-flux regime is
\begin{align}
I_{\mathrm{eff}}(\phi_+)
=& I_1\sin(\phi_+ + \theta) +
I_{1/2}\sin\!\left(\frac{\phi_+}{2}+\chi_{1/2}\right)
\nonumber\\ &
+I_2\sin(2\phi_+ + \chi_2),
\label{eq:app_comb_Ieff0}
\end{align}
with
\begin{align}
I_1 &= J_0(\xi_0)R_0,
\label{eq:app_comb_I1}\\
I_{1/2} &= \frac{i_{c2}^{(1)}}{2}J_0\!\left(\frac{\xi_0}{2}\right),
\label{eq:app_comb_Ihalf}\\
I_2 &= \frac{i_{c2}^{(2)}}{2}J_0(2\xi_0),
\label{eq:app_comb_I2}
\end{align}
and
\begin{align}
R_0 &=
\sqrt{
\left(\frac{1+\alpha}{2}\right)^2\cos^2\Psi
+
\left(\frac{1-\alpha}{2}\right)^2\sin^2\Psi
},
\label{eq:app_comb_R0}\\
\tan\theta &=
\frac{(1-\alpha)\sin\Psi}{(1+\alpha)\cos\Psi},
\label{eq:app_comb_theta}\\
\chi_{1/2} &= -\frac{\Psi}{2},
\qquad
\chi_2=-2\Psi.
\label{eq:app_comb_chi}
\end{align}

To incorporate the conventional amplitude-asymmetry diode mechanism,
we now add the induced correction derived in Section~\ref{app:analytics}. To the order
retained there, the effective force acquires the extra contribution
\begin{equation}
g_{\mathrm{ind}}(\phi_+)
=
K_{\mathrm{tot}}\,H(\phi_+),
\label{eq:app_comb_gind}
\end{equation}
where
\begin{equation}
K_{\mathrm{tot}}
=
\frac{\pi\beta_L}{2}
+
\frac{\beta_c\xi_0^2}{2(1+\beta_c^2\omega^2)},
\label{eq:app_comb_Ktot}
\end{equation}
and
\begin{align}
H(\phi_+)
=&
\frac{a^2\sin^2\Psi-b^2\cos^2\Psi}{2}\sin 2\phi_+
\nonumber\\
&-ab\sin\Psi\cos\Psi\,\cos 2\phi_+,
\label{eq:app_comb_H}
\end{align}
with
\begin{equation}
a=\frac{1+\alpha}{2},
\qquad
b=\frac{1-\alpha}{2}.
\label{eq:app_comb_ab}
\end{equation}
The full asymptotic effective CPR is therefore written as
\begin{align}
&I_{\mathrm{eff}}(\phi_+)=
I_1\sin(\phi_+ + \theta)
+ g_{\mathrm{ind}}(\phi_+)
\nonumber\\
&
+ I_{1/2}\sin\!\left(\frac{\phi_+}{2}+\chi_{1/2}\right)
+ I_2\sin(2\phi_+ + \chi_2).
\label{eq:app_comb_Ieff}
\end{align}

Under the dominance condition \eqref{eq:app_comb_dom}, the extrema of the CPR are close to those of the leading dressed first harmonic,
\begin{equation}
\phi_+^{(+)}=\frac{\pi}{2}-\theta,
\qquad
\phi_+^{(-)}=-\frac{\pi}{2}-\theta.
\label{eq:app_comb_extrema}
\end{equation}
We therefore write
\begin{equation}
I_{\mathrm{eff}}(\phi_+)
=
I_1\sin(\phi_+ + \theta)+g(\phi_+),
\label{eq:app_comb_decomp}
\end{equation}
where
\begin{align}
g(\phi_+)=& g_{\mathrm{ind}}(\phi_+)
+ I_{1/2}\sin\!\left(\frac{\phi_+}{2}+\chi_{1/2}\right)
\nonumber\\ &
+I_2\sin(2\phi_+ + \chi_2).
\label{eq:app_comb_g}
\end{align}
To first order in \(g\), the shifts of the extrema do not contribute,
and the switching currents are
\begin{align}
I_c^+ &\simeq 2\Bigl[I_1+g(\phi_+^{(+)})\Bigr],
\label{eq:app_comb_Icp}\\
|I_c^-| &\simeq 2\Bigl[I_1-g(\phi_+^{(-)})\Bigr].
\label{eq:app_comb_Icm}
\end{align}
Hence
\begin{equation}
\eta
\simeq
\frac{g(\phi_+^{(+)})+g(\phi_+^{(-)})}{2I_1}.
\label{eq:app_comb_eta_general}
\end{equation}

From Section ~\ref{app:analytics}, the fast-averaged second-harmonic function
$\langle H(\phi_+)\rangle$ summed at the unperturbed extrema
is given by Eq.~(\ref{eq:H_extrema_np}),
\begin{equation}
\bigl\langle H\bigr\rangle_{\phi_+^{(+)}}
+\bigl\langle H\bigr\rangle_{\phi_+^{(-)}}
=
\bigl[1+J_0(2\xi_0)\bigr]\,
\frac{\alpha(1-\alpha^2)\sin 2\Psi}{8R_0^2}.
\label{eq:app_comb_Hcrit}
\end{equation}
Therefore,
\begin{align}
&g_{\mathrm{ind}}(\phi_+^{(+)})+g_{\mathrm{ind}}(\phi_+^{(-)})
\nonumber\\
&
=K_{\mathrm{tot}}\bigl[1+J_0(2\xi_0)\bigr]
\frac{\alpha(1-\alpha^2)\sin 2\Psi}{8R_0^2},
\label{eq:app_comb_sum_ind}
\end{align}
and its contribution to the diode efficiency is
\begin{equation}
\eta_{\mathrm{ind}}
=
\frac{K_{\mathrm{tot}}\bigl[1+J_0(2\xi_0)\bigr]\,\alpha(1-\alpha^2)\sin 2\Psi}
{16\,J_0(\xi_0)\,R_0^3},
\label{eq:app_comb_eta_ind}
\end{equation}
in agreement with the non-perturbative result
Eq.~(\ref{eq:eta_a_nonpert_app}). If one expands
\begin{equation}
J_0(\xi_0)=1+O(\xi_0^2),
\qquad
J_0(2\xi_0)=1+O(\xi_0^2),
\end{equation}
so that $[1+J_0(2\xi_0)]/[2J_0(\xi_0)]\to 1$,
then Eq.~\eqref{eq:app_comb_eta_ind} reduces, up to the neglected mixed
terms of Eq.~\eqref{eq:app_comb_mixed}, to the Section~\ref{app:analytics} result
\begin{equation}
\eta_{\mathrm{ind}}
\simeq
\eta_L+\Delta\eta_{ac},
\label{eq:app_comb_eta_ind_A}
\end{equation}
with
\begin{align}
\eta_L
&=
\frac{\pi\beta_L\,\alpha(1-\alpha^2)\sin(2\Psi)}
{16R_0^3},
\label{eq:app_comb_etaL}\\
\Delta\eta_{ac}
&=
\frac{\alpha(1-\alpha^2)\phi_{ac}^2\sin(2\Psi)}
{4\beta_c\beta_L^2R_0^3(1+\beta_c^2\omega^2)
\left[(\omega_L^2-\omega^2)^2+\gamma^2\omega^2\right]}.
\label{eq:app_comb_Deltaeta}
\end{align}

Using the results already derived in Section \ref{app:jacobi_anger},
\begin{align}
&\sin\!\left(
\frac{\phi_+^{(+)}}{2}+\chi_{1/2}
\right)
+
\sin\!\left(
\frac{\phi_+^{(-)}}{2}+\chi_{1/2}
\right)
\nonumber\\
&\qquad
=
-\sqrt{2}\,
\sin\!\left(\frac{\Psi+\theta}{2}\right),
\label{eq:app_comb_half_sum}
\end{align}
so that
\begin{equation}
\eta_{1/2}
=
-\frac{I_{1/2}}{\sqrt{2}\,I_1}
\sin\!\left(\frac{\Psi+\theta}{2}\right).
\label{eq:app_comb_eta_half0}
\end{equation}
Substituting Eqs.~\eqref{eq:app_comb_I1} and
\eqref{eq:app_comb_Ihalf} gives
\begin{equation}
\eta_{1/2}
=
-\frac{
i_{c2}^{(1)}J_0(\xi_0/2)
}{
2\sqrt{2}\,J_0(\xi_0)\,R_0
}
\sin\!\left(\frac{\Psi+\theta}{2}\right).
\label{eq:app_comb_eta_half}
\end{equation}

Likewise,
\begin{align}
\sin(2\phi_+^{(+)}+\chi_2)
+
\sin(2\phi_+^{(-)}+\chi_2)
=
2\sin\!\bigl[2(\Psi+\theta)\bigr],
\label{eq:app_comb_second_sum}
\end{align}
and therefore
\begin{equation}
\eta_2
=
\frac{I_2}{I_1}\sin\!\bigl[2(\Psi+\theta)\bigr].
\label{eq:app_comb_eta20}
\end{equation}
Using Eqs.~\eqref{eq:app_comb_I1} and \eqref{eq:app_comb_I2},
\begin{equation}
\eta_2
=
\frac{
i_{c2}^{(2)}J_0(2\xi_0)
}{
2\,J_0(\xi_0)\,R_0
}
\sin\!\bigl[2(\Psi+\theta)\bigr].
\label{eq:app_comb_eta2}
\end{equation}

Adding the three contributions, we obtain the combined asymptotic
formula
\begin{equation}
\eta
\simeq
\eta_{\mathrm{ind}}
+
\eta_{1/2}
+
\eta_2
\label{eq:app_comb_eta_total0}
\end{equation}
or explicitly
\begin{widetext}
\begin{equation}
\eta
\simeq
\frac{K_{\mathrm{tot}}\bigl[1+J_0(2\xi_0)\bigr]\,\alpha(1-\alpha^2)\sin 2\Psi}
{16\,J_0(\xi_0)\,R_0^3}
-
\frac{
i_{c2}^{(1)}J_0(\xi_0/2)
}{
2\sqrt{2}\,J_0(\xi_0)\,R_0
}
\sin\!\left(\frac{\Psi+\theta}{2}\right)
+
\frac{
i_{c2}^{(2)}J_0(2\xi_0)
}{
2\,J_0(\xi_0)\,R_0
}
\sin\!\bigl[2(\Psi+\theta)\bigr].
\label{eq:app_comb_eta_total}
\end{equation}
\end{widetext}
Equivalently, if one expands the first term consistently to the order
used in Section~\ref{app:analytics}
\begin{widetext}
\begin{equation}
\eta
\simeq
\eta_L+\Delta\eta_{ac}
-
\frac{
i_{c2}^{(1)}J_0(\xi_0/2)
}{
2\sqrt{2}\,J_0(\xi_0)\,R_0
}
\sin\!\left(\frac{\Psi+\theta}{2}\right)
+
\frac{
i_{c2}^{(2)}J_0(2\xi_0)
}{
2\,J_0(\xi_0)\,R_0
}
\sin\!\bigl[2(\Psi+\theta)\bigr].
\label{eq:app_comb_eta_total_expanded}
\end{equation}
\end{widetext}

Eq.~\eqref{eq:app_comb_eta_total_expanded} provides a compact
summary of the three physically distinct routes to a diode effect in the present SQUID geometry. The first term,
\(\eta_L+\Delta\eta_{ac}\), is the induced contribution of a
conventional amplitude-asymmetric SQUID, which vanishes in the strict pinned-flux leading-order reduction and appears only after finite-loop feedback and dynamical back-action are included. The second and third terms are the direct contributions of the explicit \(\sin(\phi/2)\) and \(\sin 2\phi\) CPR harmonics, whose distinct Bessel dressings \(J_0(\xi_0/2)\) and \(J_0(2\xi_0)\) are responsible for the different arc spacings in the two-dimensional map \(\eta(\phi_{ac},\omega)\).

The combined asymptotic formula is valid only when the dressed first
harmonic remains dominant and the omitted mixed terms remain small.
In particular, it should not be used quantitatively near the nodal lines of \(J_0(\xi_0)\) or near flux values where \(R_0\to 0\), nor in the strong-drive regime where higher orders in \(\xi_0\) become important. In those regimes the full effective CPR must be analyzed without expanding in the subleading harmonics, and the critical currents must in general be extracted numerically from the complete equations of motion.

\section{Numerical protocol}
\label{app:numerics}

All numerical results were obtained by direct time-domain
integration of the coupled RCSJ equations, using \textsc{Matlab} with the adaptive Runge-Kutta solver \texttt{ode89} (tolerances $\mathrm{RelTol}=10^{-8}$,
$\mathrm{AbsTol}=10^{-10}$). For each operating point the
diode efficiency $\eta$ was extracted from a full current-voltage characteristic: the dimensionless bias current was swept over $i_b\in[-4,4]$ on $500$ points, and for each $i_b$ the equations of motion were integrated and the time-averaged voltage $\langle v\rangle$ evaluated over an integer number of AC cycles after discarding transients.

The switching currents $I_c^\pm$ were identified as the
smallest positive (largest negative) bias at which $|\langle v\rangle|>V_{\mathrm{th}}=0.005$. The static maps Fig.~2, the drive-amplitude cuts $\eta(\phi_{ac})$ Fig.~3, the frequency cuts $\eta(\omega)$ Fig.~4 and the AC drive-amplitude cuts for low and high frequencies Fig.~5 of the main text were all produced with this same \textsc{Matlab} framework, parallelized over the bias-current sweep. For dynamical regimes we used the typical integration parameters (minimum integration time $T_{\max}^{\mathrm{floor}}=10^4$, $70$ AC cycles, averaging over the final $50$ cycles, and at least $400$ output points per cycle). These conservative settings were chosen to resolve the weak diode signal of the conventional reference case and to stabilize the phase boundaries.

In turn, the 2D phase diagrams $\eta(\phi_{ac},\omega)$ of Figs.~6 and~7 of the main text, which require a full bias sweep at each of the $200\times 200$ points $(\phi_{ac},\omega)\in[0.01,3.0]\times[0.75,5.0]$ for
each CPR case and each value of $\beta_c$, were computed on
the TRUBA high-performance cluster of TUBITAK, using
the same integration and switching-current extraction procedure as for \textsc{Matlab} calculations.

\end{document}